\documentclass[10pt]{article}
\usepackage{fullpage}
\usepackage{amsmath}
\usepackage{amssymb}
\usepackage{amsfonts}
\usepackage{comment}
\usepackage[dvips]{epsfig}
\usepackage{color}
\usepackage{cite}

\def\##1{\underline{#1}}
\def\=#1{\underline{\underline{#1}}}

\def\+
#1{\underline{\bf #1}}
\def\*#1{\underline{\underline{\bf #1}}}

\def\r#1{(\ref{#1})}
\def\l#1{\label{#1}}
\def\c#1{\cite{#1}}

\def\le{\left(}
\def\ri{\right)}
\def\les{\left[}
\def\ris{\right]}
\def\lec{\left\{}
\def\ric{\right\}}

\def\.{\mbox{ \tiny{$^\bullet$} }}

\def\eps{\varepsilon}

\def\epso{\eps_{\scriptscriptstyle 0}}
\def\lambdao{\lambda_{\scriptscriptstyle 0}}
\def\muo{\mu_{\scriptscriptstyle 0}}

\def\ko{k_{\scriptscriptstyle 0}}

\def\ux{\hat{\#u}_x}
\def\uy{\hat{\#u}_y}
\def\uz{\hat{\#u}_z}

\def\calA{{\cal A}}
\def\calB{{\cal B}}

\def\PAmat{\les\=P_\calA\ris}
\def\PBmat{\les\=P_\calB\ris}

\begin{document}

\begin{center}

\LARGE{ {\bf  On Dyakonov--Voigt surface waves guided by the planar interface of dissipative materials
}}
\end{center}
\begin{center}
\vspace{10mm} \large

Chenzhang Zhou\\
 {\em NanoMM~---~Nanoengineered Metamaterials Group\\ Department of Engineering Science and Mechanics\\
Pennsylvania State University, University Park, PA 16802--6812, USA}
\\
 \vspace{3mm}
 Tom G. Mackay\footnote{E--mail: T.Mackay@ed.ac.uk.}\\
{\em School of Mathematics and
   Maxwell Institute for Mathematical Sciences\\
University of Edinburgh, Edinburgh EH9 3FD, UK}\\
and\\
 {\em NanoMM~---~Nanoengineered Metamaterials Group\\ Department of Engineering Science and Mechanics\\
Pennsylvania State University, University Park, PA 16802--6812,
USA}\\
 \vspace{3mm}
 Akhlesh  Lakhtakia\\
 {\em NanoMM~---~Nanoengineered Metamaterials Group\\ Department of Engineering Science and Mechanics\\
Pennsylvania State University, University Park, PA 16802--6812, USA}

\normalsize

\end{center}

\begin{center}
\vspace{15mm} {\bf Abstract}
\end{center}
Dyakonov--Voigt (DV) surface waves guided by the planar interface 
of (i) material $\calA$ which is a uniaxial dielectric material specified by a relative permittivity dyadic with eigenvalues
$\eps^s_\calA$ and $\eps^t_\calA$, and (ii) material $\calB$ which is 
  an isotropic dielectric material with relative permittivity $\eps_\calB$,
  were 
   numerically investigated by solving the corresponding canonical boundary-value problem.
The two partnering materials are 
 generally dissipative,
 with the optic axis of  material $\calA$ being inclined at the angle $\chi \in \les 0 ^\circ,  90^\circ \ris$ relative to the interface plane. No solutions of the dispersion equation for DV surface waves exist when  $\chi=90^\circ$. Also, no solutions exist  for 
$\chi \in \le 0 ^\circ,  90^\circ \ri$, when both partnering materials
 are nondissipative.
For $\chi \in \les 0 ^\circ,  90^\circ \ri$, the degree of dissipation of material $\calA$ has a profound effect on
the
phase speeds, propagation lengths, and penetration depths of the DV surface waves. For mid-range values of $\chi$, DV surface waves with negative phase velocities were found. For fixed  values
of $\eps^s_\calA$ and $\eps^t_\calA$ in the upper-half-complex plane, DV surface-wave propagation is only possible for large values of $\chi$ when $| \eps_\calB|$ is very small.

\section{Introduction}

This paper concerns the propagation of electromagnetic surface waves \c{Boardman,ESW_book,Takayama_JPCM} guided by
the planar interface of two dissimilar materials labeled $\calA$ and  $\calB$.
Both partnering materials are homogeneous and dielectric. The relative permittivity
dyadic $\=\eps_\calA$ of material $\calA$ is    uniaxial \cite{Chen}   with eigenvalues
$\eps^s_\calA$ and $\eps^t_\calA$. 
Material $\calB$ is isotropic with relative permittivity scalar $\eps_\calB$. It has been established very well, both  theoretically \c{Marchevskii,Dyakonov88} and experimentally \c{Takayama_exp,Pulsifer,Takayama_NN}, that 
this interface supports the propagation of Dyakonov surface waves, for certain ranges of values
of $\eps^{s}_\calA$, $\eps^{t}_\calA$, and $\eps_\calB$.

In contrast to
 surface-plasmon-polariton waves \c{Pitarke,Homola}, for example,  Dyakonov surface waves 
 propagate without decay  when both partnering materials are nondissipative \c{DSWreview,Walker98,Alshits,Polo_EM}. 
 Typically,  
  Dyakonov surface waves  propagate only for a small range of  directions parallel to the interface plane \c{DSWreview,Nelatury};  but  much larger ranges of
directions are possible, 
if  partnering materials   are dissipative \c{Sorni,MLieee},
magnetic \c{Crasovan_OL},  or   exhibit negative-phase-velocity plane-wave
propagation \c{Crasovan_PRB}.

As we established recently \c{MZL_PRSA}, the planar interface of materials $\calA$ and $\calB$ can support another type of surface wave, namely the Dyakonov--Voigt (DV) surface wave. There are  fundamental  differences between DV surface waves and  Dyakonov surface waves:
The amplitude of a 
 DV surface wave decays in a combined exponential--linear manner with increasing distance from the interface plane in material $\calA$, whereas 
 Dyakonov surface waves only decay in an exponential manner with inceasing distance from the interface plane.
  Furthermore, a 
 DV surface wave propagates  in only one  direction
 in each quadrant of the interface plane whereas Dyakonov surface waves propagate for a range of directions in each quadrant of the interface plane  \c{DSWreview,Walker98}.
 
 The fields of the DV surface wave in the partnering material $\calA$ 
have similar characteristics to the fields associated with a singular form of 
 planewave propagation called Voigt-wave propagation \c{Voigt}.
 The existence of a Voigt wave for an unbounded anisotropic material 
 is a consequence of the corresponding propagation matrix being non-diagonalizable \c{Panch,Grech, Ranganath,Gerardin}.
The feature that distinguishes  a Voigt wave from a conventional plane wave  \cite{Chen,EAB} is that the Voigt wave's amplitude depends on
the product of an
exponential function of the propagation distance and a linear function of the
propagation distance, the latter being absent for a conventional plane wave.

DV surface-wave propagation has been established  \c{MZL_PRSA} when both partnering materials and nondissipative and
  the optic axis of material $\calA$ lies wholly in the interface plane.
 In the following sections,  DV surface-wave propagation is investigated for the case where the partnering materials are generally dissipative and the optic axis of material $\calA$ is inclined at the angle $\chi$ relative to the interface plane.
 As we demonstrate, the incorporation of dissipation and inclination of the optic axis
 of material $\calA$
  has profound effects upon the phase speeds, propagation lengths, and penetration depths of the DV surface waves.

In the notation adopted,
 the permittivity and permeability of free space are denoted by $\epso$ and $\muo$, respectively. 
 The free-space wavelength is written as $\lambdao = 2 \pi / \ko$ with
 $\ko = \omega \sqrt{\epso \muo}$ being
 the free-space wavenumber 
 and  $\omega$ being  the angular frequency. The
real and imaginary parts of  complex-valued quantities
are delivered by the operators $\mbox{Re} \lec \. \ric$ 
and $\mbox{Im} \lec \. \ric$, respectively, and $i = \sqrt{-1}$. 
Single underlining denotes a 3-vector 
and     $\lec \ux, \uy, \uz \ric$ is
the triad of unit vectors aligned with the Cartesian axes.
Square brackets enclose matrixes and column vectors. The superscript ${}^T$ denotes the transpose. The complex conjugate is denoted by an asterisk.

\section{Theory}

\subsection{Preliminaries}

\begin{figure}[!htb]
\centering
\includegraphics[width=8.6cm]{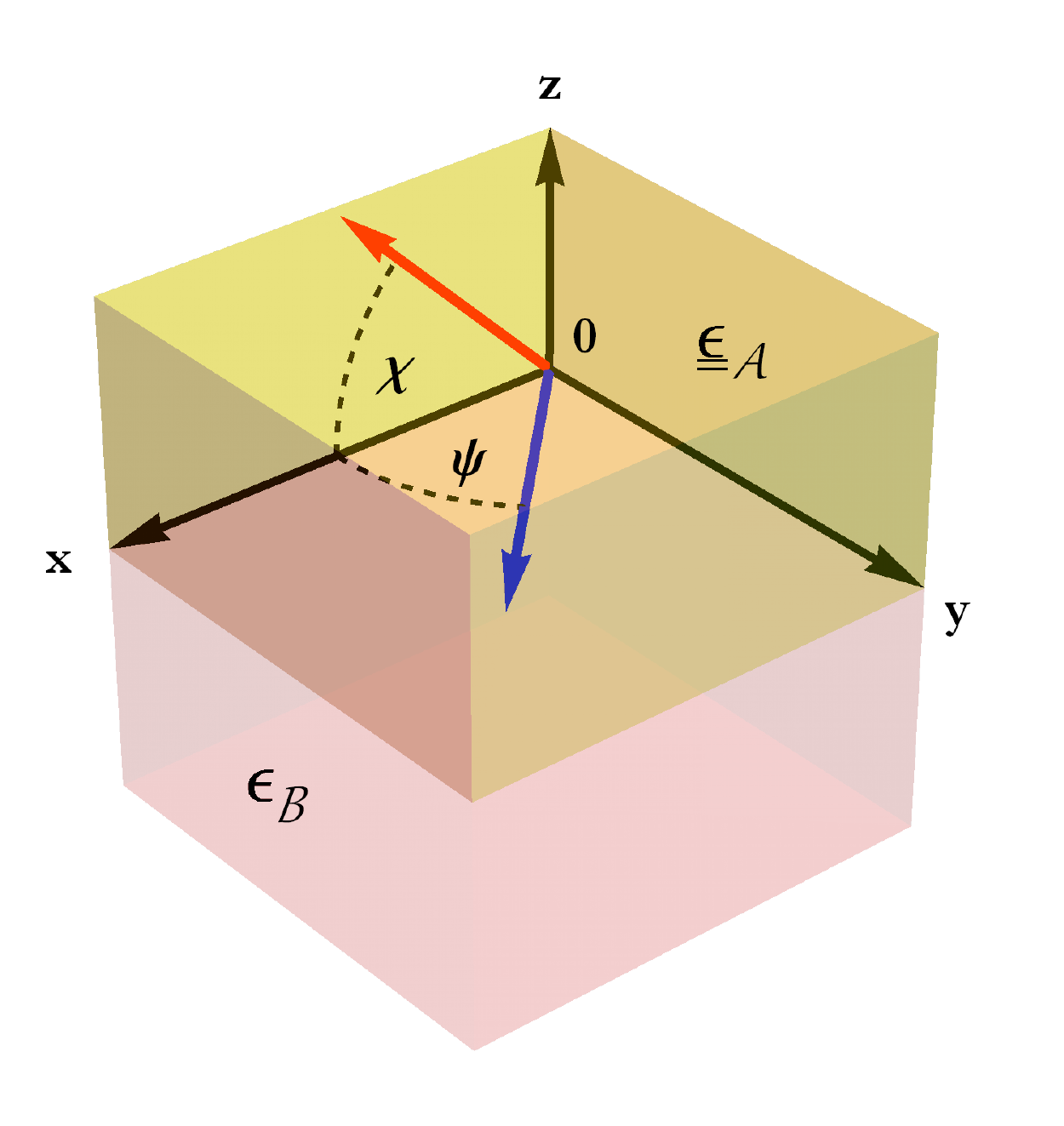} \caption{Schematic representation of the canonical boundary-value problem, in which 
material $\calA$ occupies the half-space $z>0$ and material $\calB$
 the half-space $z<0$. 
The optic axis of material $\calA$ lies at the angle
$\chi$ to the $x$ axis in the $xz$ plane.
The surface wave propagates
parallel to the interface  plane  $z=0$ at the angle $\psi$ relative to the $x$ axis.
 } \label{fig1}
\end{figure}

In the canonical boundary-value problem for DV surface-wave propagation shown
in  Fig.~\ref{fig1},
material $\calA$   occupies the half-space $z>0$ and material $\calB$   the half-space $z<0$.
The relative permittivity dyadic  of material $\calA$ is given as
\begin{equation}\={\eps}_\calA = \=S_y(\chi) \.
\les {\eps}_\calA^t \ux\, \ux +  {\eps}_\calA^s \le \uy\, \uy + \uz \, \uz \ri \ris  \.
 \=S_y^T(\chi), \end{equation}
wherein the  rotation dyadic
\begin{equation}
\=S_y(\chi) = \uy\,\uy + \le \ux \, \ux  + \uz \, \uz \ri \cos \chi
+\le \uz \, \ux  - \ux \, \uz \ri \sin \chi.
\end{equation}
Thus, the optic axis of material $\calA$ lies wholly in the  $xz$ plane
at an angle $\chi$ with respect to the $x$ axis. 
 The relative permittivity dyadic of material $\calB$
 is specified as $\=\eps_\calB = \eps_\calB \=I$, where $\=I=\ux\, \ux +
  \uy\, \uy + \uz \, \uz$ is the 3$\times$3 identity dyadic  \cite{Chen}.
The relative permittivity parameters $\eps^s_\calA$, $\eps^t_\calA$, and 
$\eps_\calB$ are complex valued, in general.  

The theory for DV
surface-wave propagation in the case of $\chi = 0^\circ$ for nondissipative partnering materials is provided in the predecessor paper \c{MZL_PRSA}. In this section, the theory is extended to cover the  $0^\circ < \chi \leq 90^\circ $ regime for dissipative partnering
materials.

The electromagnetic field phasors for surface-wave propagation 
are expressed everywhere as
\cite{ESW_book} 
\begin{equation} \label{planewave}
\left.\begin{array}{l}
 \#E (\#r)=  \les e_x(z)\ux + e_y(z)\uy+e_z(z)\uz \ris \, \\ \hspace{12mm} \times \exp\les i q \le x \cos \psi + y \sin \psi \ri \ris \\[4pt]
 \#H  (\#r)=  \les h_x(z)\ux + h_y(z)\uy+h_z(z)\uz \ris \,
 \\ \hspace{12mm} \times
 \exp\les i q \le x \cos \psi + y \sin \psi \ri \ris 
 \end{array}\right\}\,,  \:\:\: -\infty < z < + \infty,
\end{equation}
with ${q}$ being the surface  wavenumber. The angle $\psi\in\left[0,2\pi\right)$
specifies
 the direction of propagation in the $xy$ plane, relative to  the $x$ axis.
 The phasor representations~\r{planewave}, when combined with
the source-free Faraday and Amp\'ere--Maxwell equations,  deliver 
 the 4$\times$4 matrix ordinary differential
equations \c{Berreman}
\begin{equation}
\label{MODE_A}
\frac{d}{dz}\les\#f  (z)\ris= \left\{
\begin{array}{l}
i \PAmat\.\les\#f  (z)\ris\,,  \qquad   z>0 \vspace{8pt} \\
i \PBmat\.\les\#f  (z)\ris\,,  \qquad   z<0
\end{array} , \right.
\end{equation}
wherein
the  column 4-vector
\begin{equation}
\les\#f (z)\ris= 
\les
\begin{array}{c}%
e_ x(z), \quad
e_y(z),\quad
h_x(z),\quad
h_y(z)
\end{array}
\ris^T,
\label{f-def}
\end{equation}
and  the 4$\times$4 propagation matrixes  $\les\=P_\ell\ris $, $\ell\in\lec\calA,\calB\ric$, are determined by  $\=\eps_\ell$.
The $x$-directed and $y$-directed components of the phasors
are algebraically connected to their   $z$-directed components  \c{EAB}.

\subsection{Half-space $z>0$}

The  4$\times$4 propagation matrix  $\les\=P_\calA\ris $ is given as
\begin{equation}
\les\=P_\calA\ris=  \les   
\begin{array}{cccc}
\frac{\beta }{ \gamma }&0& 
\frac{q^2 \sin 2\psi}{2 \omega \epso \gamma } & 
 \frac{k_o^2 \gamma- \nu_c}{\omega \epso \gamma}   \\
\frac{\beta \tan \psi }{ \gamma}&0& 
 \frac{\nu_s -k_o^2 \gamma }{\omega \epso \gamma}&
\frac{-q^2 \sin 2\psi}{2 \omega \epso \gamma}
\\
 \frac{-q^2 \sin 2\psi}{2 \omega \muo} & 
\frac{-\ko^2 \eps^s_\calA+ \nu_c }{\omega \muo}  &0&0\\
\frac{\ko^2 \eps^s_\calA \eps^t_\calA- \gamma \nu_s }{\omega  \muo  \gamma }  &
\frac{q^2 \sin 2\psi}{2 \omega \muo}&
- \frac{\beta \tan \psi}{ \gamma }&
\frac{\beta }{ \gamma }
\end{array}
\ris, 
\end{equation}
wherein the generally complex-valued parameters 
\begin{equation}
\left.
\begin{array}{l}
\nu_c = q^2  \cos^2 \psi \vspace{0pt} \\
\nu_s = q^2  \sin^2 \psi \vspace{0pt} \\
\beta=  q \le \eps^s_\calA- \eps^t_\calA  \ri \sin \chi \cos \chi  \cos \psi \vspace{0pt} \\
\gamma=  \eps^s_\calA \cos^2 \chi + \eps^t_\calA \sin^2 \chi 
\end{array}
\right\}.
\end{equation}
The    $z$-directed  components of the field phasors are
\begin{equation}
\left.
\begin{array}{l}
e_z(z) = \displaystyle{\frac{1}{\gamma} \left\{\frac{q \les h_ x(z) \sin \psi - h_ y(z)  \cos \psi  \ris}{\omega \epso } \right.}  \vspace{4pt}  \\ \hspace{12mm}
+ \displaystyle{ e_x(z) \le \eps^s_\calA - \eps^t_\calA  \ri \sin \chi \cos \chi \Bigg\}}  \vspace{8pt} \\
h_ z(z) = \displaystyle{\frac{q \les e_ y(z) \cos \psi - e_ x(z)  \sin \psi  \ris}{\omega \muo }}
\end{array}
\right\}\,,\qquad z > 0\,.
\end{equation}

\subsubsection{Dyakonov surface wave}

In order to deal with DV surface waves, it is necessary  to first consider
 the  case of Dyakonov surface waves for which  $\les\=P_\calA\ris$ has four eigenvalues, 
each with algebraic multiplicity $1$ and geometric multiplicity $1$. 
These eigenvalues are
\begin{equation} \l{a_decay_const}
\left.
\begin{array}{l}
\alpha_{\calA a} = i \sqrt{ q^2 - \ko^2 \eps_\calA^s} \vspace{8pt}\\
\alpha_{\calA b} = - i \sqrt{ q^2 - \ko^2 \eps_\calA^s} \vspace{8pt}\\
\alpha_{\calA c} = \displaystyle{\frac{ \beta + i  \sqrt{ \eps^s_\calA \les  \nu_s \cos^2 \chi \le  \eps^s_\calA -  \eps^t_\calA \ri + q^2 \eps^t_\calA \ris -  \gamma \eps^s_\calA \eps^t_\calA k_o^2 } }{\gamma} }
 \vspace{8pt}\\
\alpha_{\calA d} = \displaystyle{\frac{  \beta - i  \sqrt{ \eps^s_\calA \les  \nu_s \cos^2 \chi \le  \eps^s_\calA -  \eps^t_\calA \ri + q^2 \eps^t_\calA \ris -  \gamma \eps^s_\calA \eps^t_\calA k_o^2 }}{\gamma}}
\end{array}
\right\}\,.
\end{equation}
Only those  eigenvalues which have positive  imaginary parts are relevant for  surface-wave propagation.
Clearly, only one of 
 $ \alpha_{\calA a} $ 
 and $ \alpha_{\calA b} $ can have a positive imaginary part. Furthermore, it is assumed that  
 only one of 
 $ \alpha_{\calA c} $ 
 and $ \alpha_{\calA d} $ can have a positive  imaginary part. Therefore the two eigenvalues that are chosen \cite{ESW_book} for our surface-wave analysis are 
 \begin{equation} \l{alphaA1}
 \begin{array}{l}
 \alpha_{\calA 1} = \left\{ 
 \begin{array}{lr}
 \alpha_{\calA a} & \quad \mbox{if} \quad \mbox{Im} \lec 
 \alpha_{\calA a} \ric > 0 \vspace{0pt}\\
 \alpha_{\calA b} & \quad  \mbox{otherwise}
 \end{array}
 \right. 
 \end{array}
 \end{equation}
 and
  \begin{equation} \l{alphaA2}
 \begin{array}{l}
 \alpha_{\calA 2} = \left\{ 
 \begin{array}{lr}
 \alpha_{\calA c} & \quad \mbox{if} \quad \mbox{Im} \lec 
 \alpha_{\calA c} \ric > 0 \vspace{0pt}\\
 \alpha_{\calA d} & \quad  \mbox{otherwise}
 \end{array}
 \right. 
 \end{array}.
 \end{equation}

\subsubsection{Dyakonov--Voigt surface wave}
In the case of DV surface-wave propagation,
we have $\alpha_{\calA 1} = \alpha_{\calA 2} = \alpha_\calA$. Thus,
 $\les\=P_\calA\ris$ has only two eigenvalues, 
each with algebraic multiplicity $2$ and geometric multiplicity $1$. 
There are are four possible values of $q$ that result in $\alpha_{\calA 1} = \alpha_{\calA 2}$, namely
\begin{equation}
q = 
\left\{
\begin{array}{l}
\displaystyle{
\frac{\ko \sqrt{\eps^s_\calA} \cos \chi \le \cos \psi \pm i \sin \chi \sin \psi \ri}{1 - \cos^2 \chi \sin^2 \psi}} \vspace{8pt} \\
\displaystyle{
- \frac{\ko \sqrt{\eps^s_\calA} \cos \chi \le \cos \psi \pm i \sin \chi \sin \psi \ri}{1 - \cos^2 \chi \sin^2 \psi}}
\end{array}
\right. ,
\end{equation}
 with 
the  correct value of $q$ for DV surface-wave propagation being the one that yields 
$\mbox{Im} \lec \alpha_{\calA} \ric > 0 $ \cite{ESW_book}.

Explicit expressions for a corresponding eigenvector $\#v_\calA$ satisfying
\begin{equation}
\le \les\=P_\calA\ris - \alpha_{\calA } \=I \ri \. \#v_{\calA } = \#0,
\end{equation}
and a corresponding generalized eigenvector $\#w_\calA$ satisfying
 \c{Boyce}
\begin{equation}
\le \les\=P_\calA\ris - \alpha_{\calA } \=I \ri \. \#w_{\calA } = \#v_{\calA },
\end{equation}
can be derived, but these expressions are too cumbersome to be reproduced here.

Thus, the general solution of Eq.~\r{MODE_A}${}_1$  representing DV surface waves that decay as $z \to +\infty$ is given as
\begin{equation} \l{DV_gen_sol}
\les\#f  (z)\ris = \les C_{\calA 1}  \#v_{\calA }  + C_{\calA 2} \le i  z \, \#v_{\calA}   + \#w_{\calA} \ri  \ris \exp \le i \alpha_{\calA } z \ri\,,\quad z > 0\,.
\end{equation}
The complex-valued constants $C_{\calA 1}$ and $C_{\calA 2}$ herein are fixed 
by applying  boundary conditions at $z=0$.

\subsection{Half-space $z<0$}

The  4$\times$4 matrix $\PBmat$   is given as \cite{ESW_book}
\begin{eqnarray} \nonumber
\les\=P_\calB\ris= 
\les   
\begin{array}{cccc}
0&0& \displaystyle{ \frac{\tau}{\omega \epso \eps_\calB}} & 
\displaystyle{\frac{\ko^2 \eps_\calB- \nu_c }{\omega \epso \eps_\calB} } \vspace{8pt} \\
0&0& \displaystyle{\frac{ \nu_s -\ko^2 \eps_\calB }{\omega \epso \eps_\calB} }&
\displaystyle{ -\frac{\tau}{\omega \epso \eps_\calB}} \vspace{8pt}
\\
\displaystyle{ -\frac{\tau}{\omega \muo}} & 
\displaystyle{\frac{\nu_c -\ko^2 \eps_\calB}{\omega \muo} } &0&0\vspace{8pt} \\
\displaystyle{\frac{\ko^2 \eps_\calB- \nu_s}{\omega \muo} } &
\displaystyle{ \frac{\tau}{\omega \muo}}&0&0
\end{array}
\ris, \\ &&
\end{eqnarray}
wherein the generally complex-valued parameter
\begin{equation}
\tau = q^2 \cos \psi \sin \psi.
\end{equation}
The $z$-directed components of the  phasors are given by 
\begin{equation}
\left.
\begin{array}{l}
e_ z(z) = \displaystyle{\frac{q \les h_ x(z) \sin \psi - h_ y(z)  \cos \psi  \ris}{\omega \epso \eps_\calB}} \vspace{8pt} \\
h_ z(z) = \displaystyle{\frac{q \les e_ y(z) \cos \psi - e_ x(z)  \sin \psi  \ris}{\omega \muo }}
\end{array}
\right\}\,,\qquad z < 0\,.
\end{equation}

Matrix $\les\=P_\calB\ris$ has two distinct eigenvalues,
each with algebraic multiplicity $2$ and geometric multiplicity $2$. These are denoted as
 $\pm \alpha_{\calB}$, with
\begin{equation} \l{b_decay_const}
\alpha_{\calB} =- i \sqrt{q^2 - \ko^2 \eps_\calB} .
\end{equation}
 The sign of the square root  in Eq.~\r{b_decay_const} must be such that 
 $\mbox{Im} \lec \alpha_{\calB} \ric < 0 $, for surface-wave propagation.
A pair of independent  eigenvectors of 
matrix $\les\=P_\calB\ris$  corresponding to the eigenvalue $\alpha_{\calB}$ 
are
\begin{equation}
\left.
\begin{array}{l}
\#v_{\calB 1} = 
\les 
\displaystyle{1 - \frac{\nu_c}{\ko^2 \eps_\calB}}, \quad
\displaystyle{- \frac{\tau}{\ko^2 \eps_\calB}}, \quad 0, \quad
 \displaystyle{\frac{\alpha_\calB}{\omega \muo}}
\ris^T
\vspace{8pt}
\\
\#v_{\calB 2} = 
\les 
\displaystyle{ \frac{\tau}{\ko^2 \eps_\calB}}, \quad
\displaystyle{ \frac{\nu_s}{\ko^2 \eps_\calB} - 1}, \quad
 \displaystyle{\frac{\alpha_\calB}{\omega \muo}, \quad 0}\:\:
\ris^T
\end{array}
\right\}.
\end{equation}
Thus, the general solution of Eq.~\r{MODE_A}${}_2$   for surface 
waves that decay as $z \to -\infty$ is provided as
\begin{equation}
\label{2.22-AL}
\les\#f  (z)\ris = \le C_{\calB 1}  \#v_{\calB 1}  +  C_{\calB 2} \#v_{\calB 2} \ri \exp \le i \alpha_{\calB} z \ri 
\,,\qquad z < 0\,,
\end{equation}
with the complex-valued constants $C_{\calB 1}$ and $C_{\calB 2}$ being fixed by applying boundary conditions at $z=0$.

\subsection{Application of boundary conditions}

 The continuity of the tangential  components of the electric and magnetic field
 phasors across the interface plane $z=0$ imposes
 four conditions that are represented compactly as
 \begin{equation}
 \label{2.23-AL}
 \les\#f(0^+)\ris=  \les\#f(0^-)\ris
 \,.
 \end{equation}
 The combination of
  Eqs.~ \r{DV_gen_sol} and \r{2.22-AL}, along with Eq.~\r{2.23-AL},
   yields
\begin{equation}
\les \=M \ris \. \les \:
 C_{\calA 1}, \quad
  C_{\calA 2}, \quad
   C_{\calB 1}, \quad
    C_{\calB 2} \:
 \ris^T =  \les \:
 0, \quad
  0, \quad
   0, \quad
    0 \:
 \ris^T,
\end{equation}
wherein the 4$\times$4 characteristic matrix $\les \=M \ris$ must be singular for  surface-wave propagation \c{ESW_book}.
The dispersion equation 
\begin{equation}
\l{dispersion_eq}
\left\vert \les \=M \ris\right\vert = 0,
 \end{equation} whose  explicit representation  is too cumbersome for reproduction here, can be numerically solved for $q$, by the Newton--Raphson method \c{N-R} for example .

\section{Numerical studies}

In order to characterize DV surface-wave propagation supported by dissipative materials, 
 we now  explore numerical solutions to the canonical boundary-value problem
that satisfy the dispersion equation \r{dispersion_eq}.

In the special case
considered previously \c{MZL_PRSA},
 wherein the optic axis of material $\calA$ lies in the interface plane (i.e., $\chi = 0^\circ$) and the materials $\calA$ and $\calB$ are nondissipative (and inactive), certain constraints on the relative permittivity parameters of materials $\calA$ and $\calB$ can be established. However, 
 the general case 
 ($ 0^\circ \leq \chi \leq 90^\circ$ and $\eps^{s}_\calA\in \mathbb{C}$, $\eps^{t}_\calA\in \mathbb{C}$,  and $\eps_\calB \in \mathbb{C}$) considered here is much less amenable to analysis and such
constraints are not forthcoming. 

Two general observations should be noted before embarking on a presentation of numerical results:
\begin{itemize} 
\item[(i)] Only null solutions are obtained for
  $\chi = 90^\circ$,  which is therefore disregarded henceforth. 
  \item[(ii)] No solutions are found  for $ \chi\in (0^\circ ,90^\circ)$ when  ${\rm Im}\lec\eps^{s}_\calA\ric=0$, ${\rm Im}\lec\eps^{t}_\calA\ric=0$,  and ${\rm Im}\lec\eps_\calB  \ric=0$.  
 \end{itemize}
 Therefore, the exhibition  of loss (or gain) by at least one
 of the two partnering materials is a prerequisite for 
DV surface-wave propagation when the 
optic axis of material $\calA$ does not lie wholly in the interface plane.

\begin{figure}[!htb]
\centering
\includegraphics[width=4.3cm]{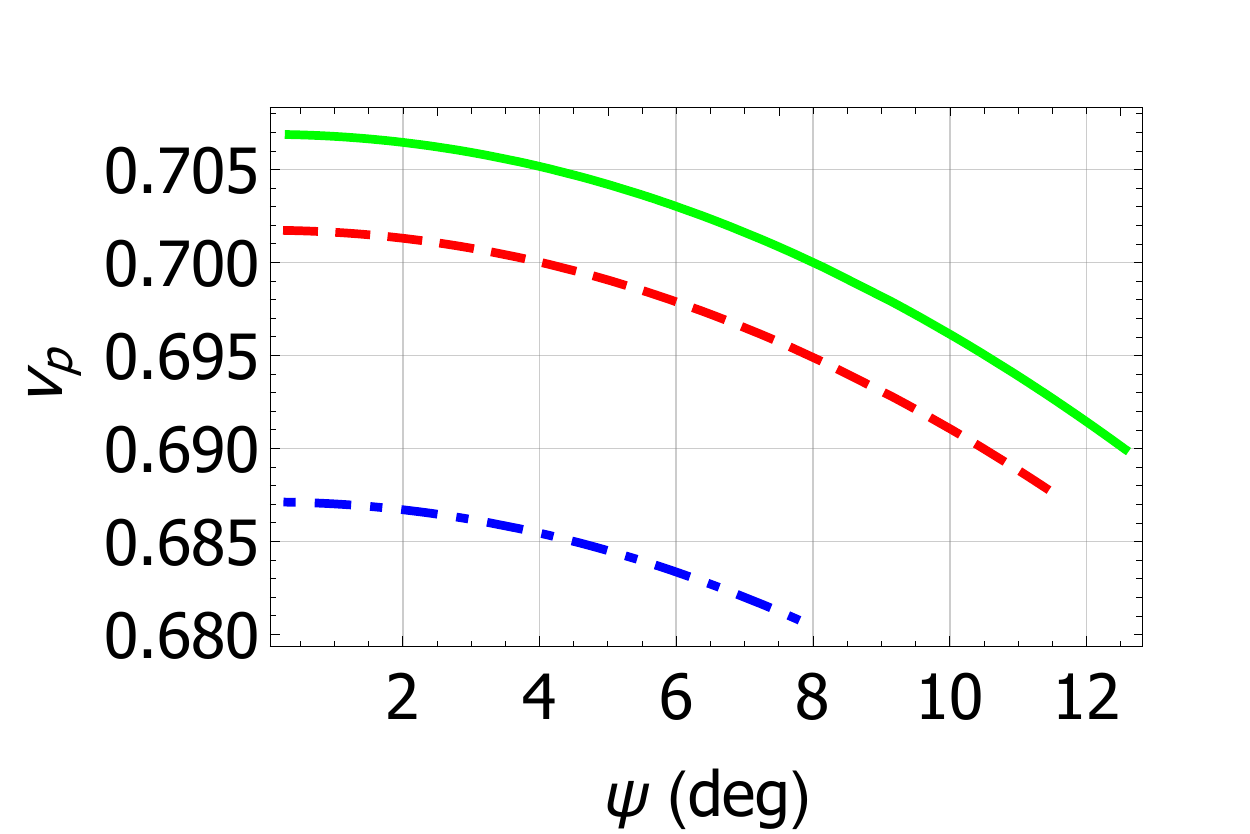}
\includegraphics[width=4.3cm]{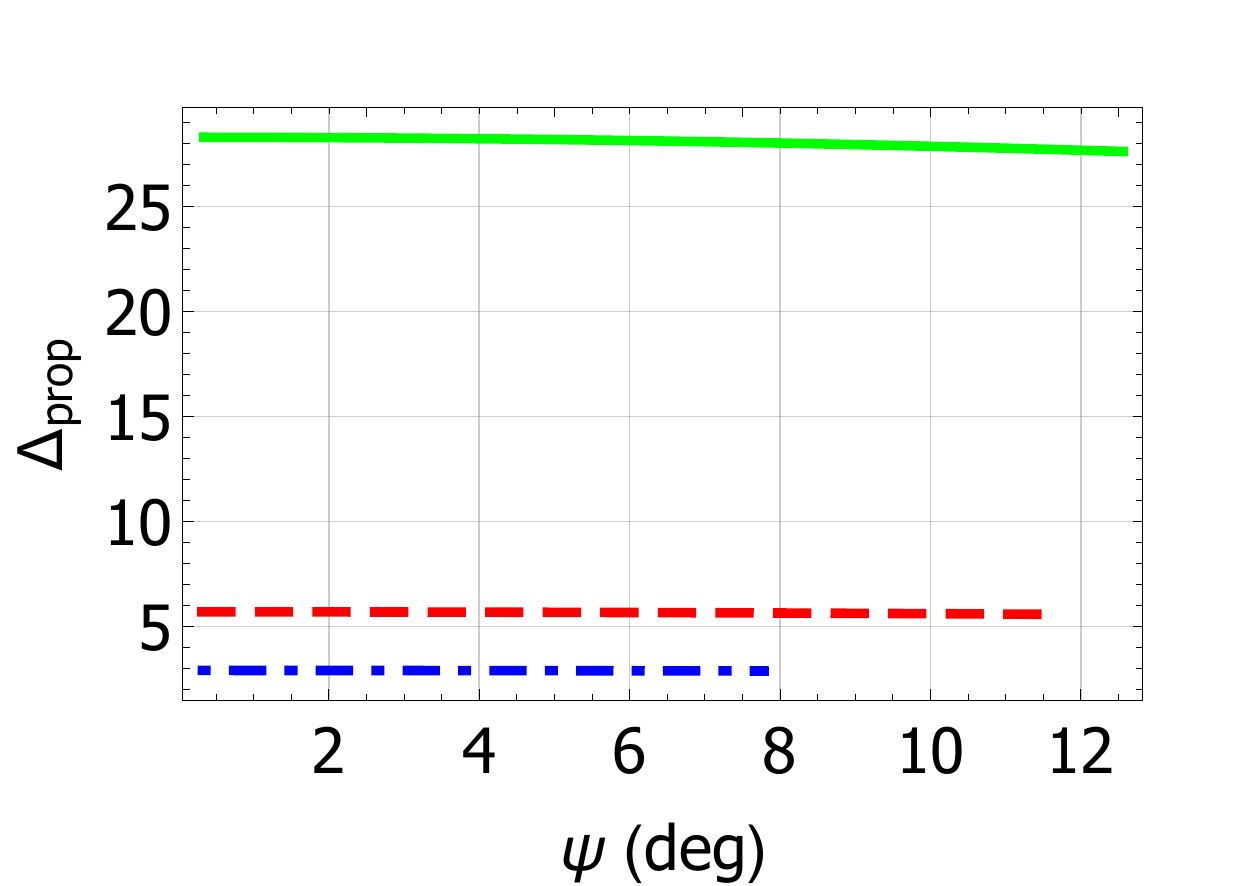}\\
\includegraphics[width=4.3cm]{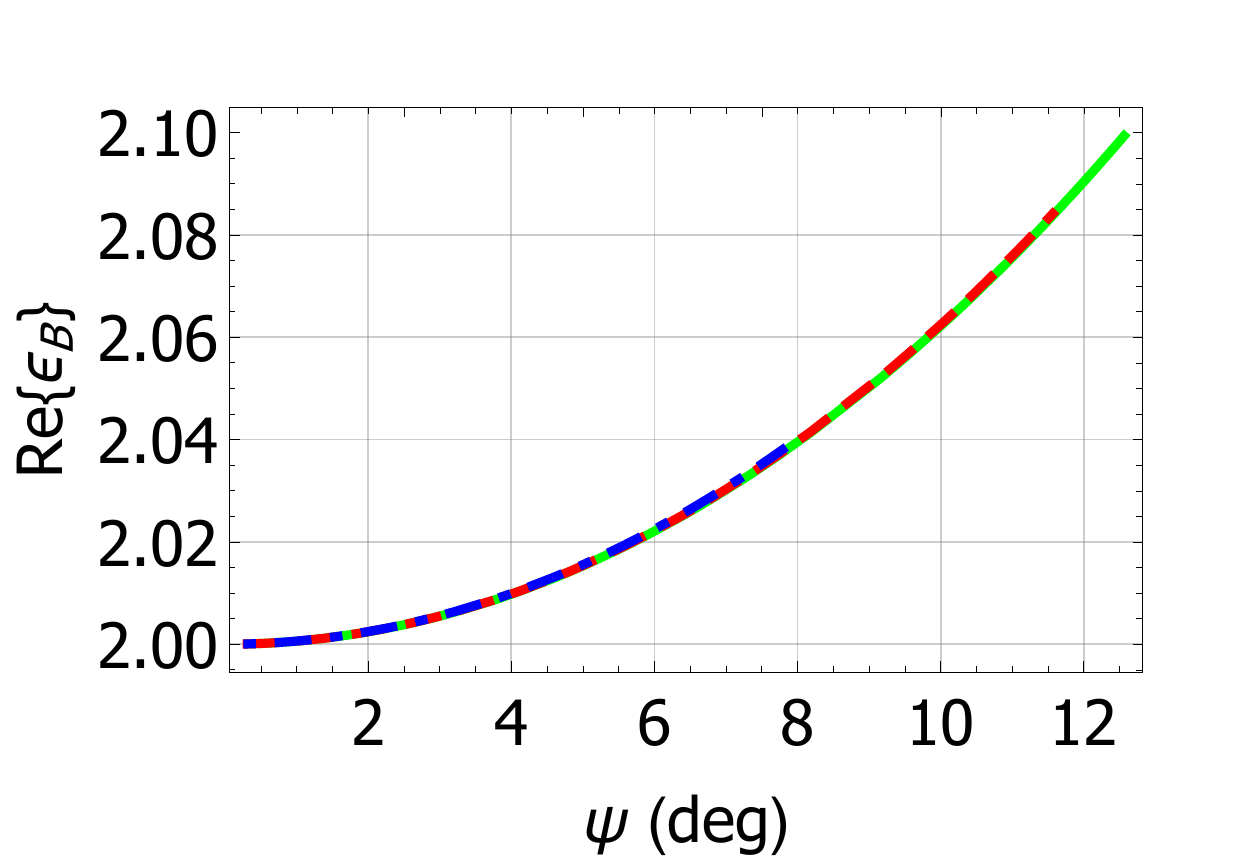}
\includegraphics[width=4.3cm]{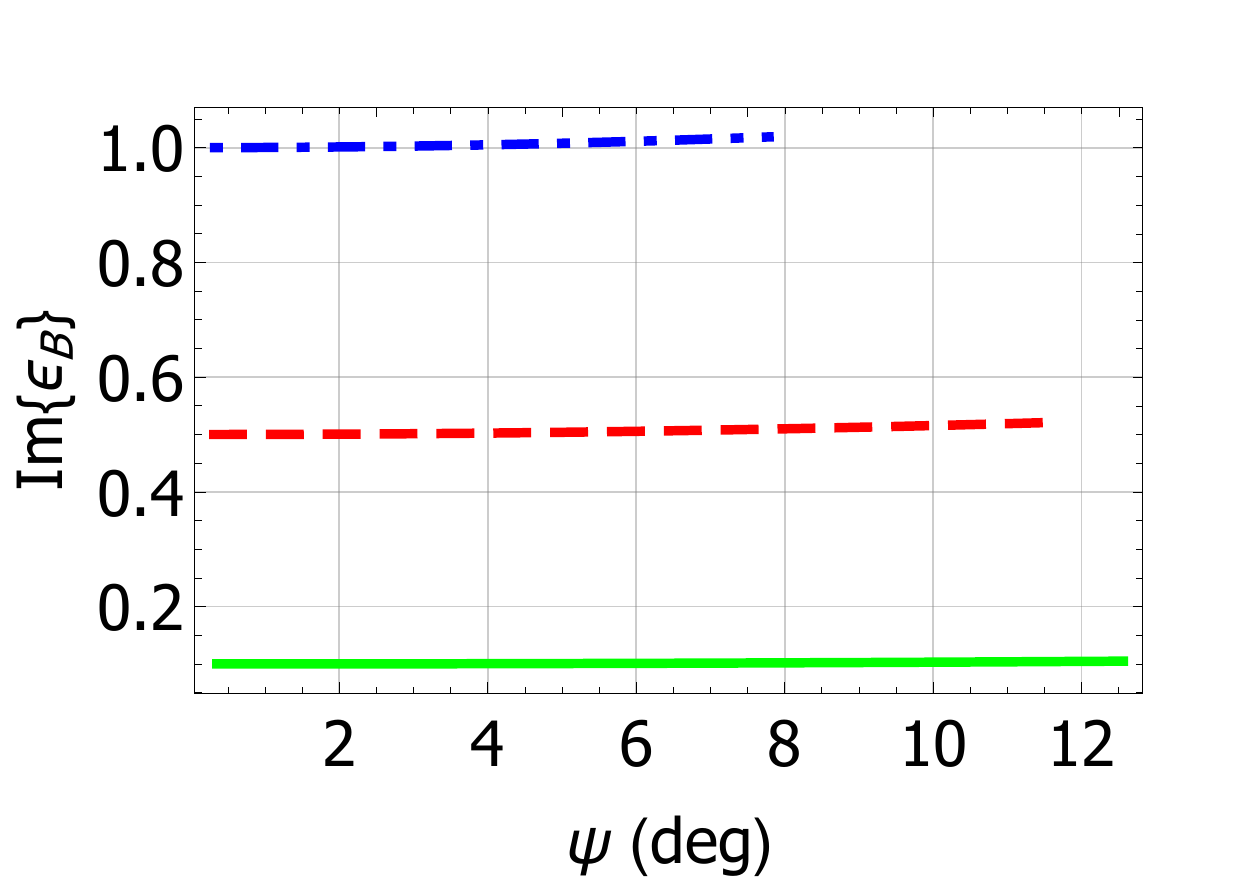}\\
\includegraphics[width=4.3cm]{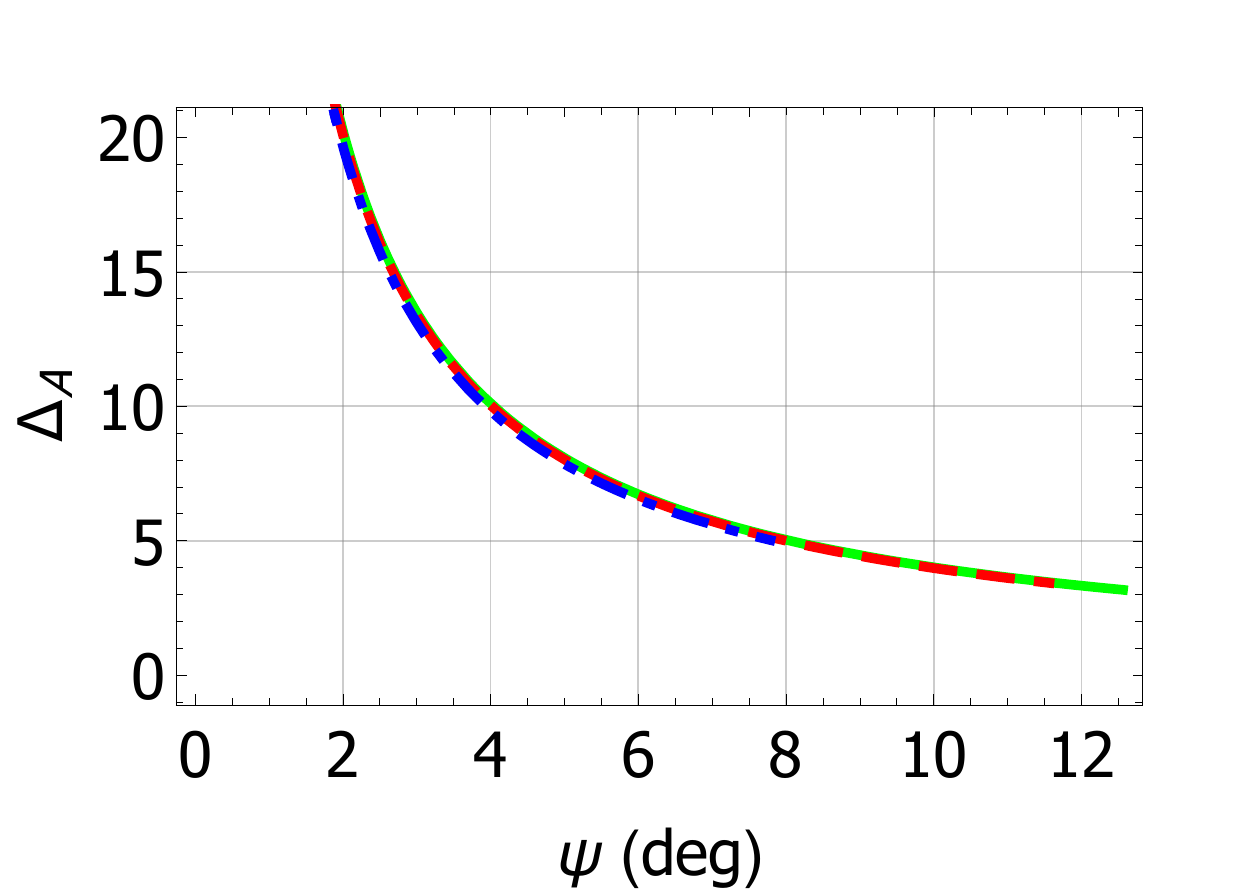}
\includegraphics[width=4.3cm]{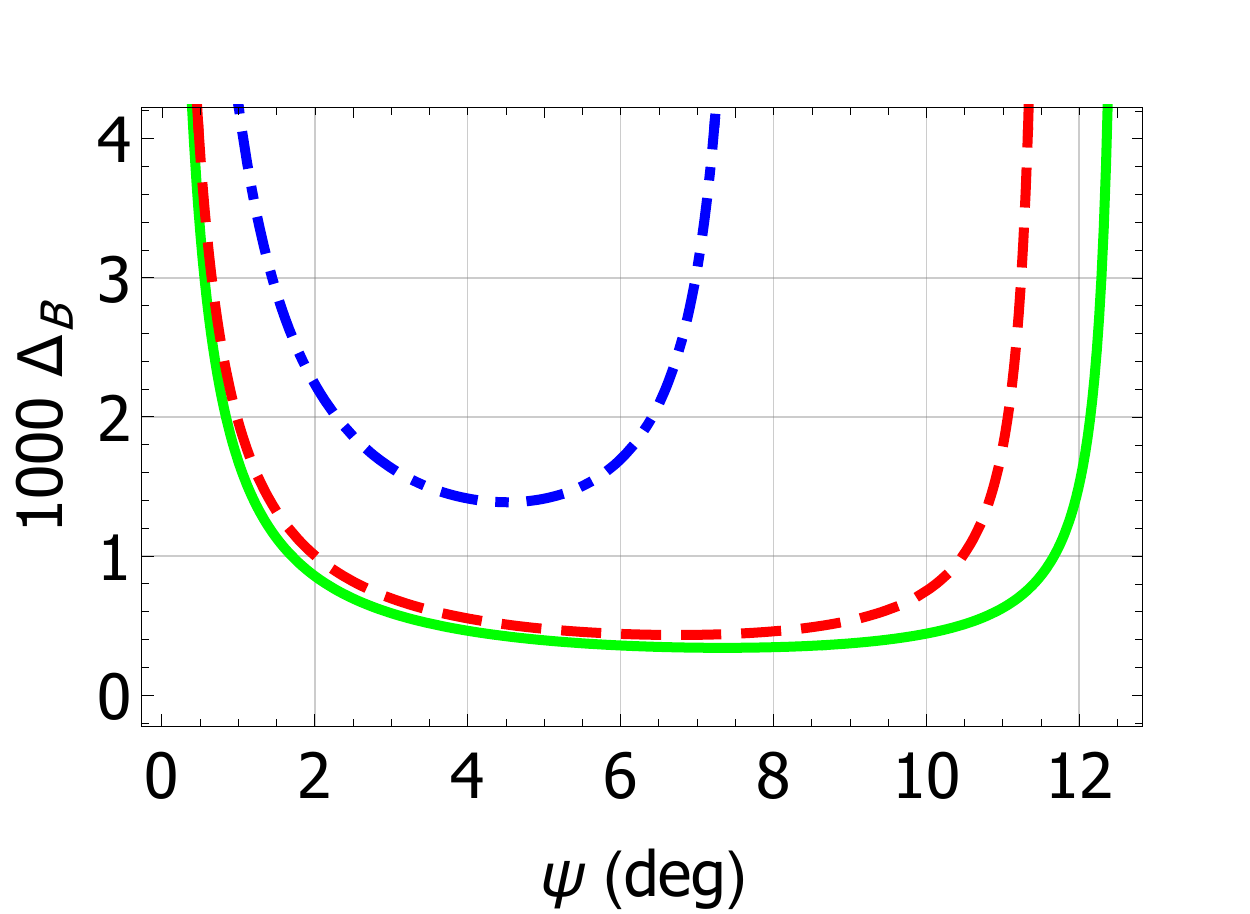}
 \caption{Real and imaginary parts of relative permittivity $\eps_\calB$,
 normalized phase speed $v_{\text{p}}$, normalized propagation length $\Delta_{\text{prop}}$, 
 normalized penetration depths $\Delta_\calA$ and 
 $\Delta_\calB$ plotted versus propagation angle $\psi$, for $\chi= 0^\circ$, $\eps^s_\calA = 2 + \delta i $, and $\eps^t_\calA = 6.2 + 2 \delta i $, when 
  $\delta = 0.1 $ (green solid curves), $\delta = 0.5 $ (red dashed  curves), and $\delta = 1 $ (blue dot-dashed curves).
 } \label{fig2}
\end{figure}

\subsection{$\chi = 0^\circ$ \label{sec3A}}
Let us begin our presentation with the case $\chi = 0^\circ$. We fixed
$\eps^s_\calA = 2 + \delta i $ and  $\eps^t_\calA = 6.2 + 2 \delta i $ for calculations. The corresponding  real and imaginary parts of  $\eps_\calB$ that satisfies the dispersion  equation \r{dispersion_eq} are plotted versus the propagation angle $\psi$ in Fig.~\ref{fig2} for $\delta \in \lec 0.1, 0.5, 1 \ric$.

Solutions only exist for relatively small  ranges of $\psi$, and these 
$\psi$-ranges shrink as $\delta$ increases. Thus, 
solutions exist for $0^\circ < \psi < 12.5^\circ$ when $\delta = 0.1$, but only for $0^\circ < \psi < 6.8^\circ$ when
$\delta = 1$. The real part of $\eps_\calB$ increases 
monotonically as $\psi$ increases, 
taking the value of $\eps^s_\calA$ at $\psi = 0^\circ$; and $\mbox{Re} \lec \eps_\calB \ric$ is very nearly independent of $\delta$.
On the other hand, the imaginary part of $\eps_B$ is very nearly independent of $\psi$, but  $\mbox{Im} \lec \eps_\calB \ric$ increases substantially as $\delta$ increases with $\mbox{Im} \lec \eps_\calB \ric \approx \mbox{Im} \lec \eps^s_\calA \ric$.

Plots of the normalized  phase speed
\begin{equation}
v_{\text{p}} = \frac{\ko }{\mbox{Re} \lec q \ric},
\end{equation}
and the normalized propagation length
\begin{equation} \l{prop_length}
\Delta_{\text{prop}} = \frac{\ko }{\mbox{Im} \lec q \ric}
\end{equation}
versus $\psi$
are also provided in Fig.~\ref{fig2}. Both $v_{\text{p}}$ and $\Delta_{\text{prop}}$ are dimensionless quantities.
The phase speed decreases monotonically as $\psi$ increases, for all values of $\delta$, with $v_{\text{p}}$ being greatest for the smallest value of $\delta$.
On the other hand, the propagation length is almost independent of $\psi$, with
$\Delta_{\text{prop}}$ being greatest  for the smallest value of $\delta$.

\begin{figure}[!htb]
\centering
\includegraphics[width=4.3cm]{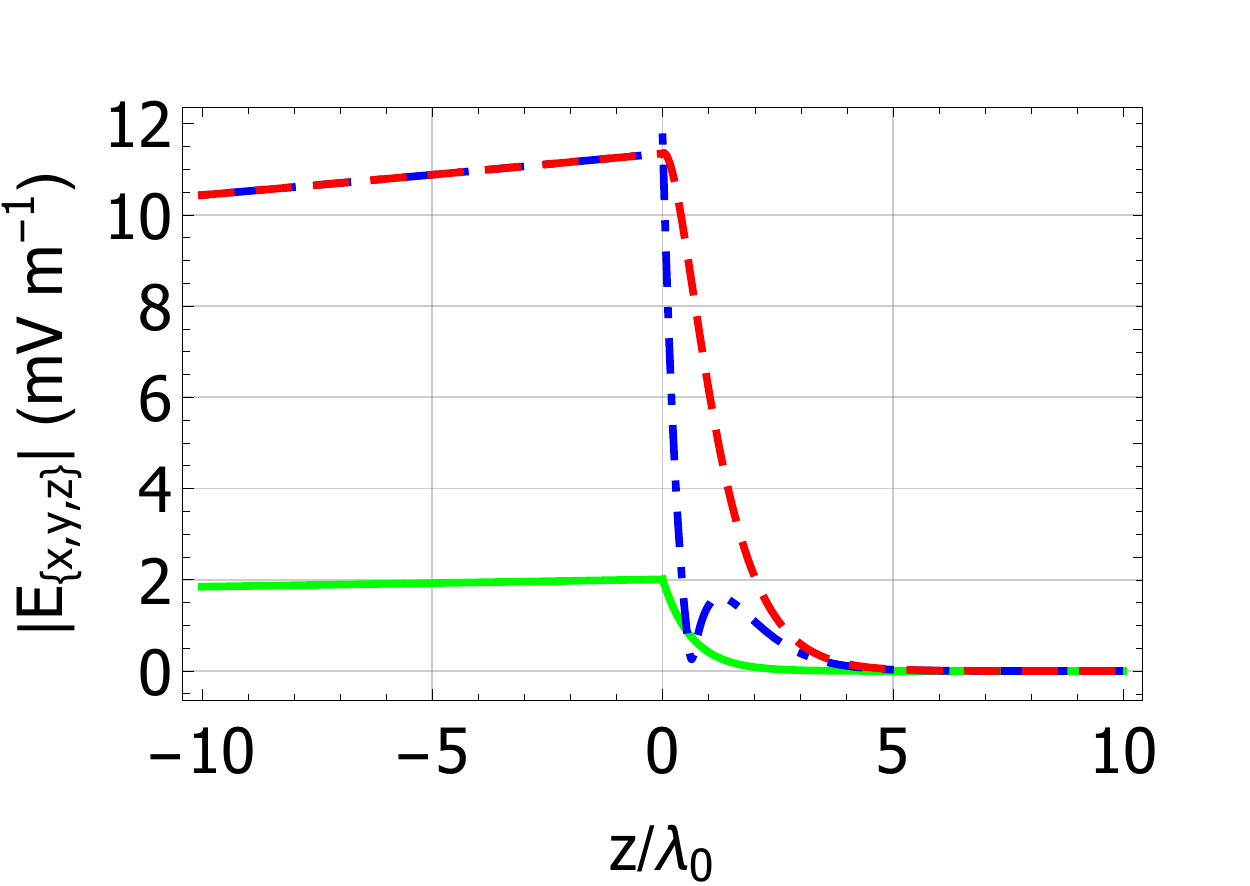}
\includegraphics[width=4.3cm]{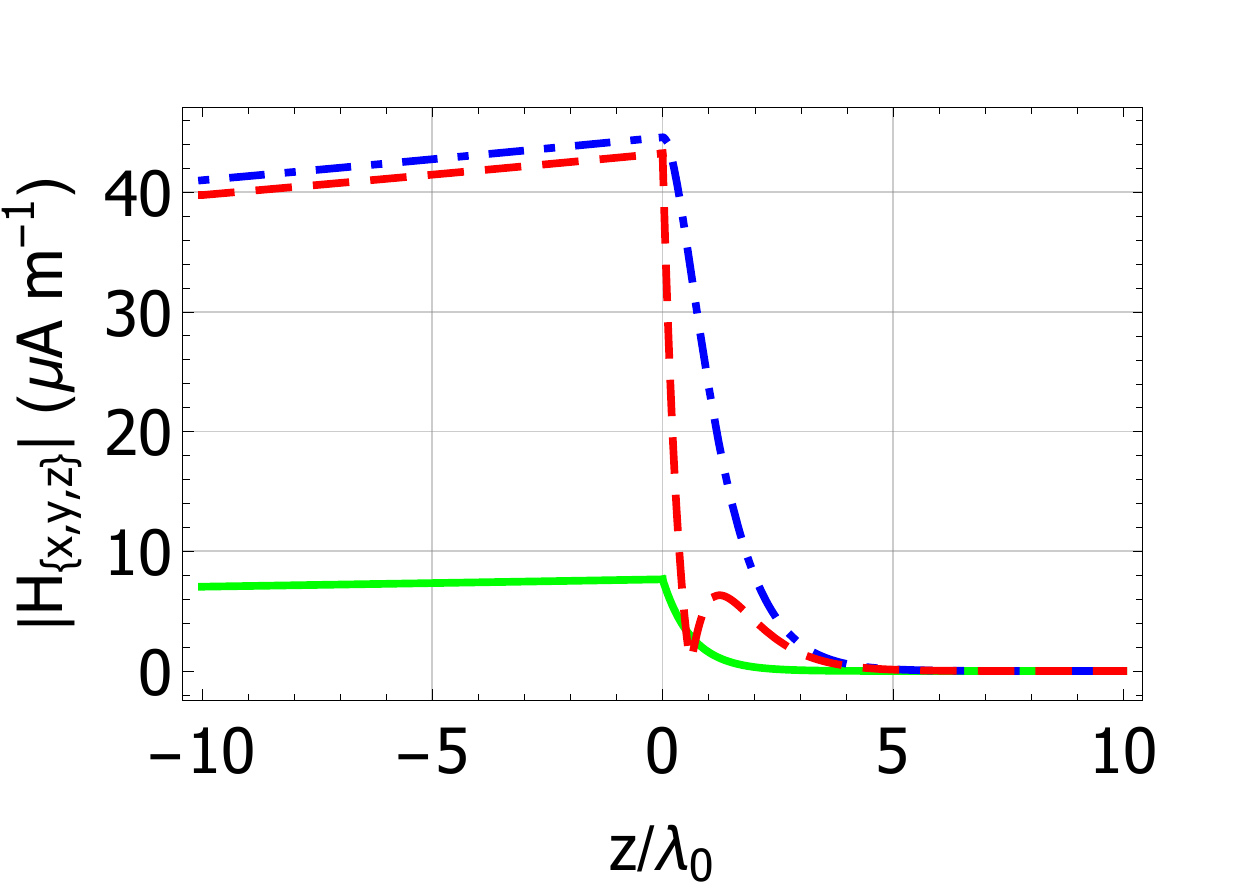}\\
\includegraphics[width=4.3cm]{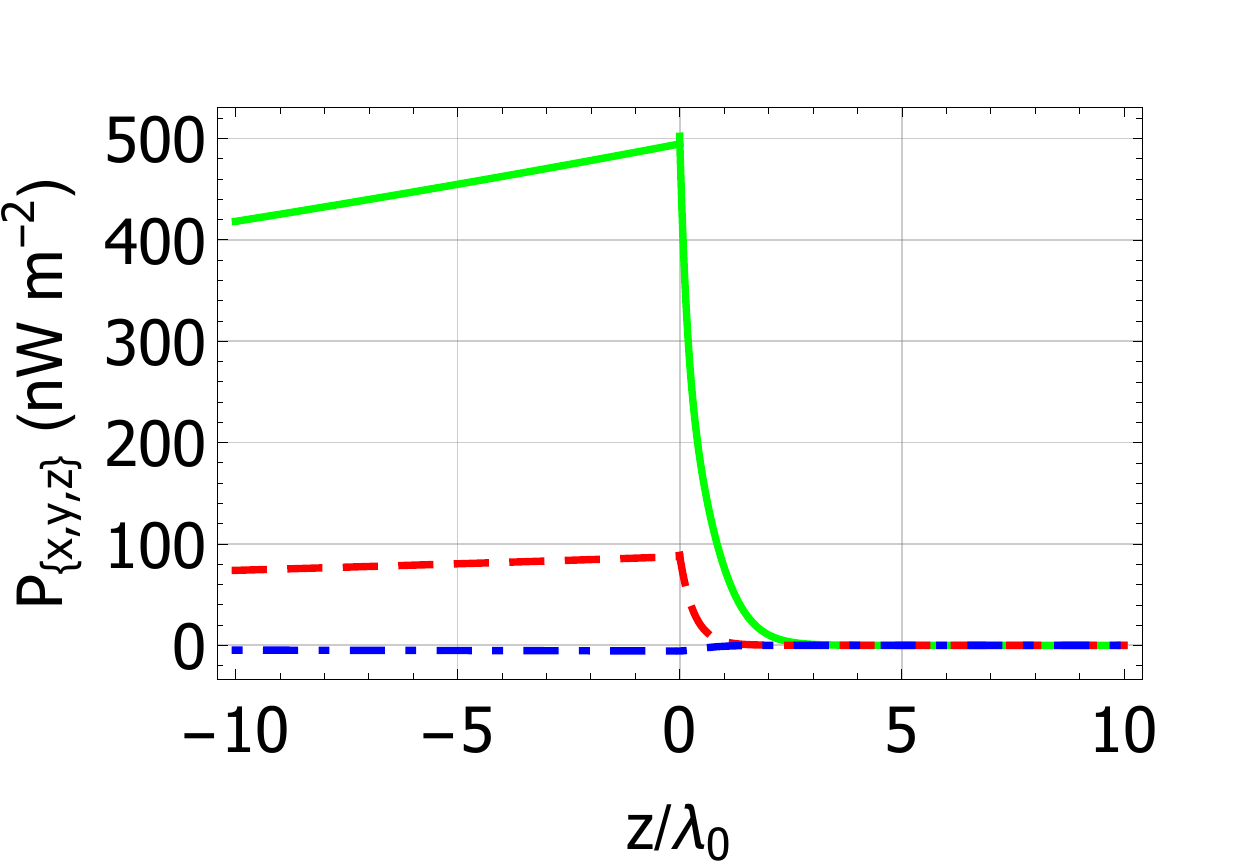}
 \caption{
$ | \#E (z \uz) \. \#n |$, $| \#H (z\uz) \. \#n |$, and 
$\#P (z\uz) \. \#n$  plotted versus $z/\lambdao$, 
 for 
  $\eps^s_\calA = 2 + 0.5 i $, $\eps^t_\calA = 6.2+i$,
 $\eps_\calB = 2.062 + 0.516i$, $\chi = 0^\circ$, and $\psi = 10^\circ$,
 with  $C_{\mathcal{B}1} = 1$~V~m${}^{-1}$, for $\#n = \ux$ 
 (green solid curves),  $\#n = \uy$ (red dashed curves), and $\#n = \uz$
 (blue dot-dashed curves).
 } \label{fig3}
\end{figure}

Also presented in Fig.~\ref{fig2} are  the
 corresponding plots of
\begin{equation} \l{pen_depths}
\left.
\begin{array}{l}
\Delta_\calA = \displaystyle{\frac{\ko}{\mbox{Im} \lec \alpha_\calA \ric}} \vspace{8pt}\\
\Delta_\calB = \displaystyle{- \frac{\ko}{\mbox{Im} \lec \alpha_\calB \ric}}
\end{array}
\right\},
\end{equation}
which represent the normalized penetration depths \cite{ESW_book} in the partnering materials $\calA$ and $\calB$,
respectively. 
Both $\Delta_\calA$ and  $\Delta_\calB$ are dimensionless quantities.
 Let us note that $\Delta_\calA$ is
very nearly independent of $\delta$; furthermore,  $\Delta_\calA$ increases monotonically as $\psi$ decreases, becoming unbounded as $\psi$ approaches $0^\circ$. In contrast, for each value of $\delta$,
 $\Delta_\calB$ has a  minimum, with the minimum value of $\Delta_\calB$ being larger for larger $\delta$. Also, $\Delta_\calB$ becomes unbounded as $\psi$ approaches either of its two extreme values.

Further light is shed by the
spatial profiles of the magnitudes of the Cartesian components of the electric and magnetic field phasors  in Fig.~\ref{fig3} for  $\delta = 0.5$ and $\psi = 10^\circ$. 
The dispersion  equation \r{dispersion_eq} then yields   $\eps_\calB = 2.062 + 0.516i$.
The  magnitudes of the components of the electric and magnetic field phasors   decay as
  the distance $\vert{z}\vert$ from the interface plane increases. The rate of decay is much faster in material $\calA$ than 
in material $\calB$.
Furthermore, beyond a short distance (approximately 1.5 $\lambdao$)  from the interface plane, 
the decay in material $\calA$ appears to exponential, from which
it may be inferred
  that  the linear term in Eq.~\r{DV_gen_sol} 
  is dominated by the exponentially decaying terms.
  
The localization of the DV  surface waves is also revealed by profiles of 
  the Cartesian components  of the time-averaged Poynting vector
\begin{equation}
\underline{P}  (\#r) = \frac{1}{2} \mbox{Re} \lec \, \underline{E}  (\#r)  \times 
\underline{H}^*  (\#r) \, \ric
\end{equation}
presented
in Fig.~\ref{fig3}.
 These profiles show that
energy flow is concentrated in directions parallel to the interface plane $z=0$.

\subsection{$\chi\in (0^\circ ,90^\circ)$ \label{sec3B}}
Next we turn to $\chi\in (0^\circ ,90^\circ)$. As in Sec.~3.\ref{sec3A},
we fix $\eps^s_\calA = 2 + \delta i $ and  $\eps^t_\calA = 6.2 + 2 \delta i $ but vary
$\delta \in \lec 0.1, 0.5, 1 \ric$.

\subsubsection{$\chi = 45^\circ$ \label{sec3B1}}

\begin{figure}[!htb]
\centering
\includegraphics[width=4.3cm]{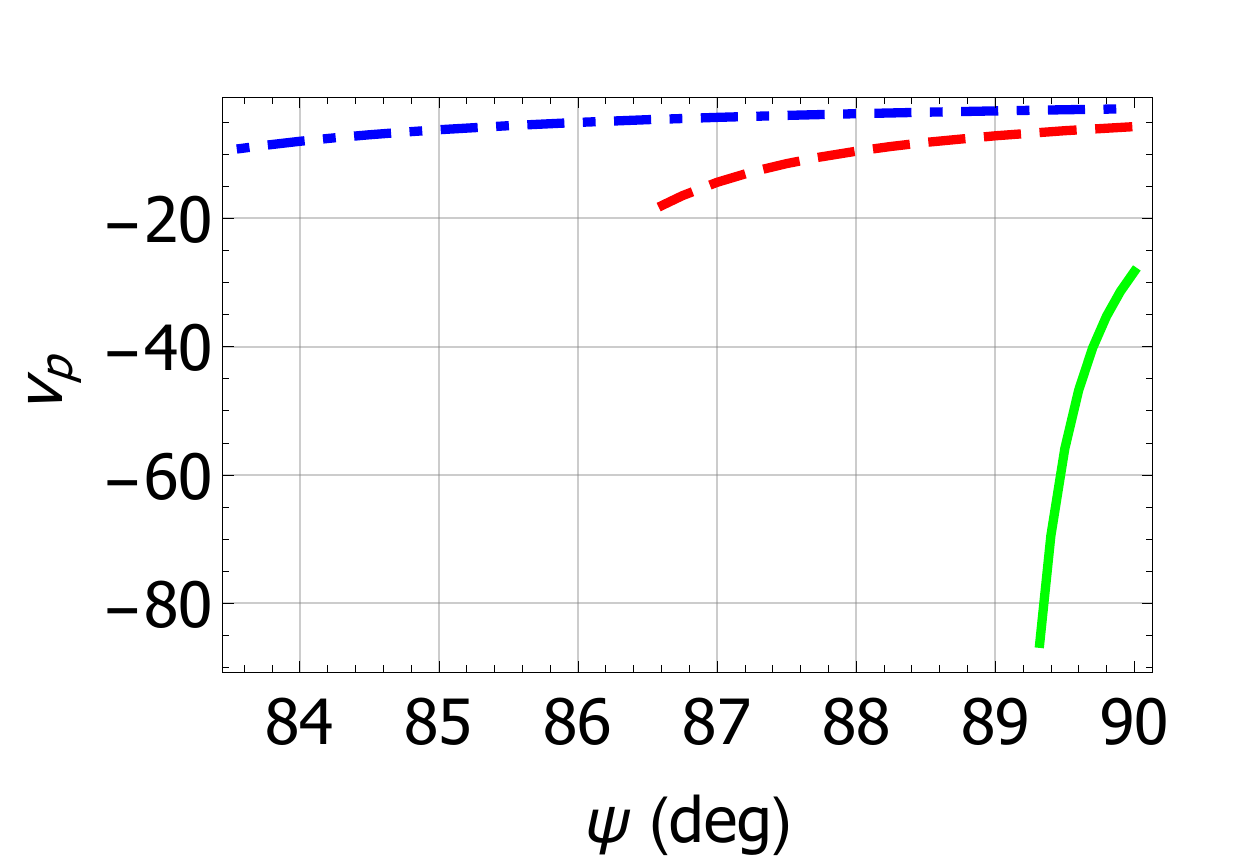}
\includegraphics[width=4.3cm]{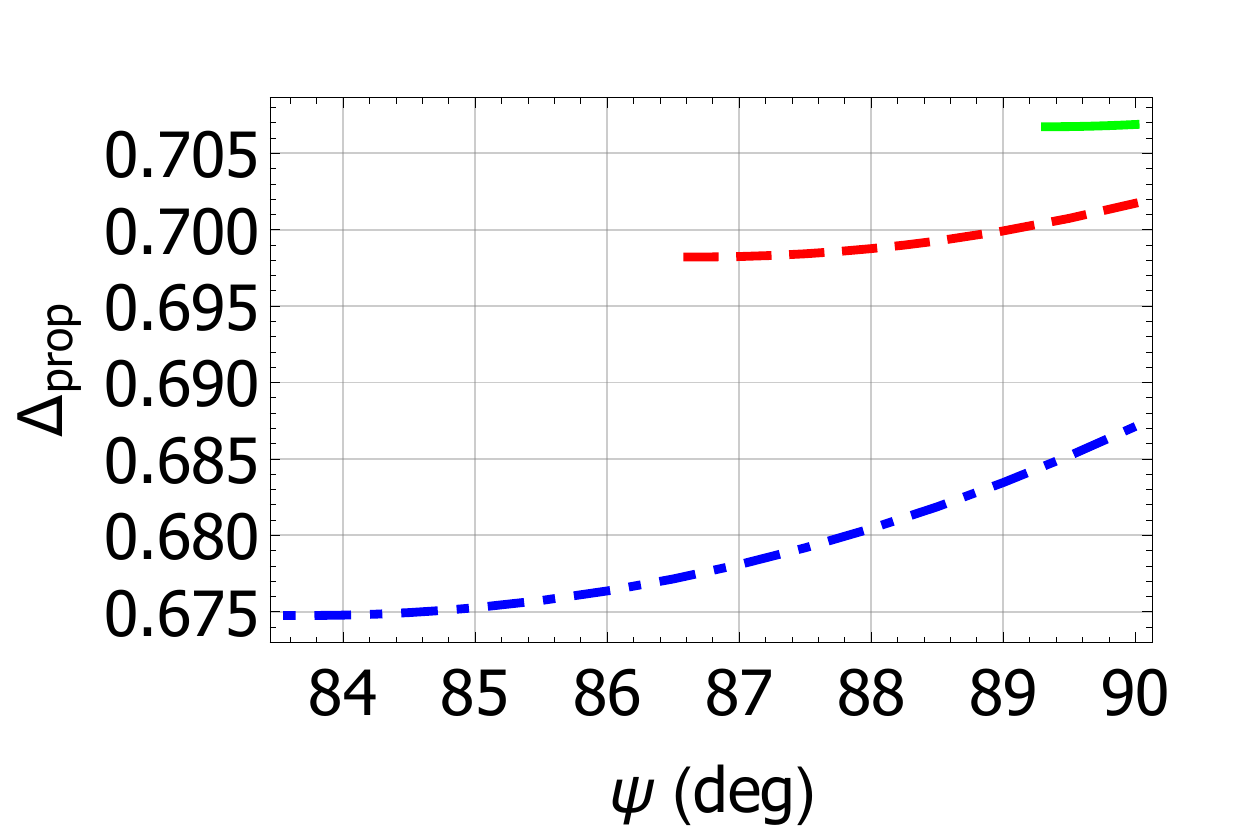}\\
\includegraphics[width=4.3cm]{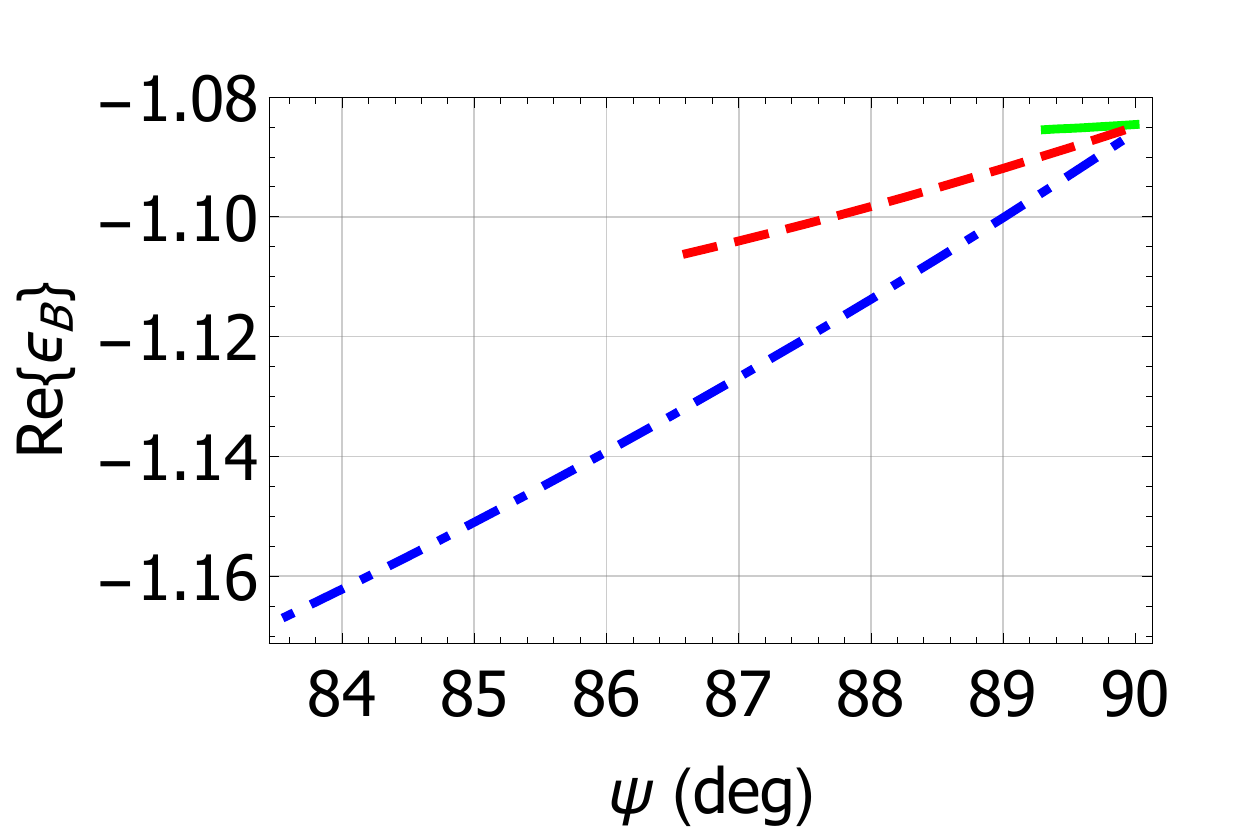}
\includegraphics[width=4.3cm]{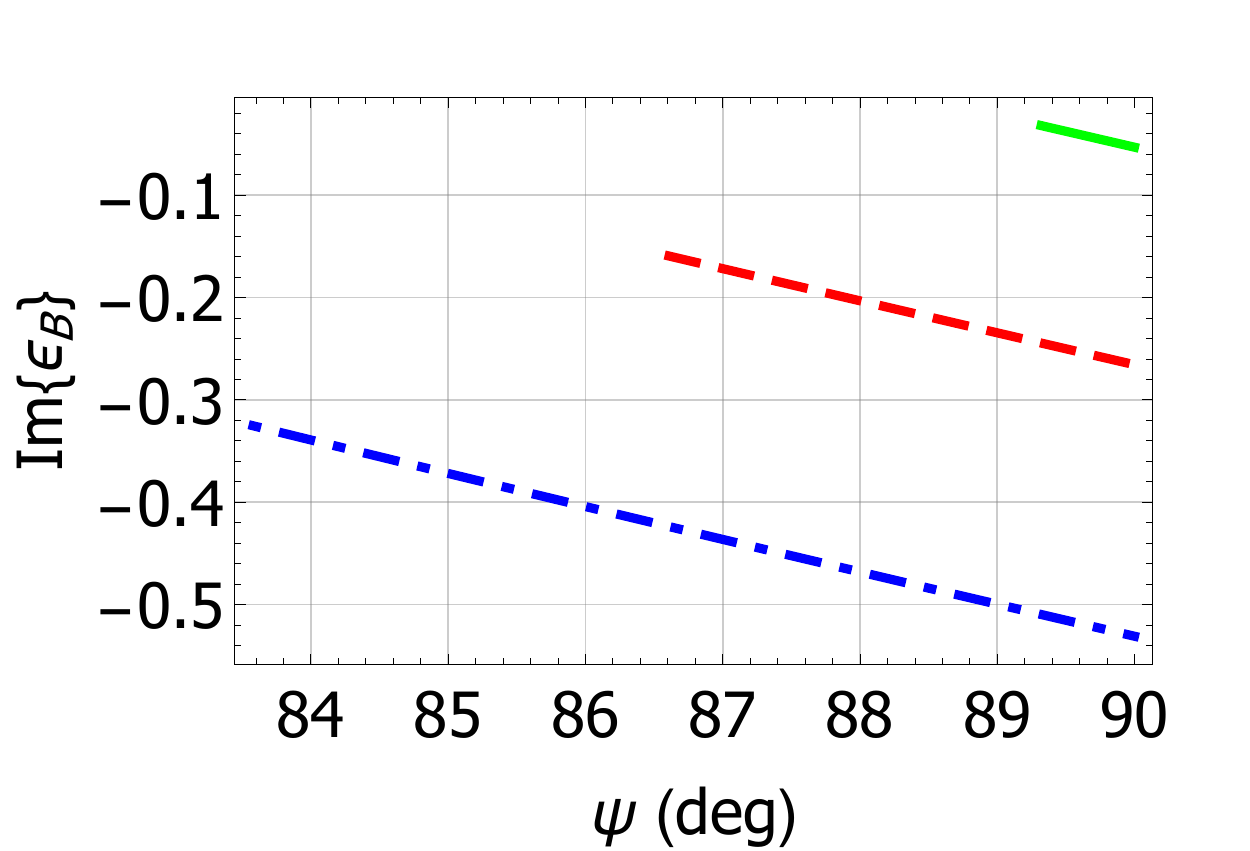}\\
\includegraphics[width=4.3cm]{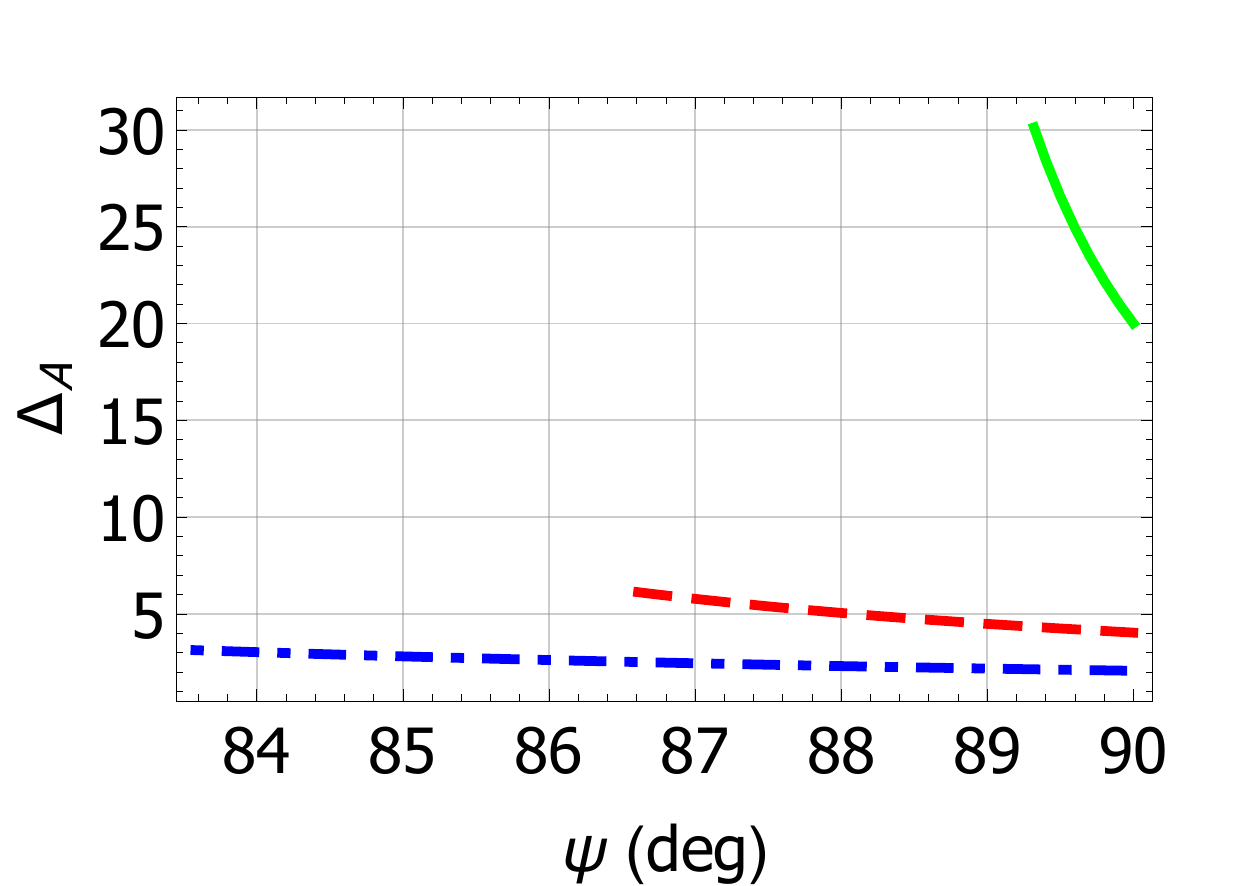}
\includegraphics[width=4.3cm]{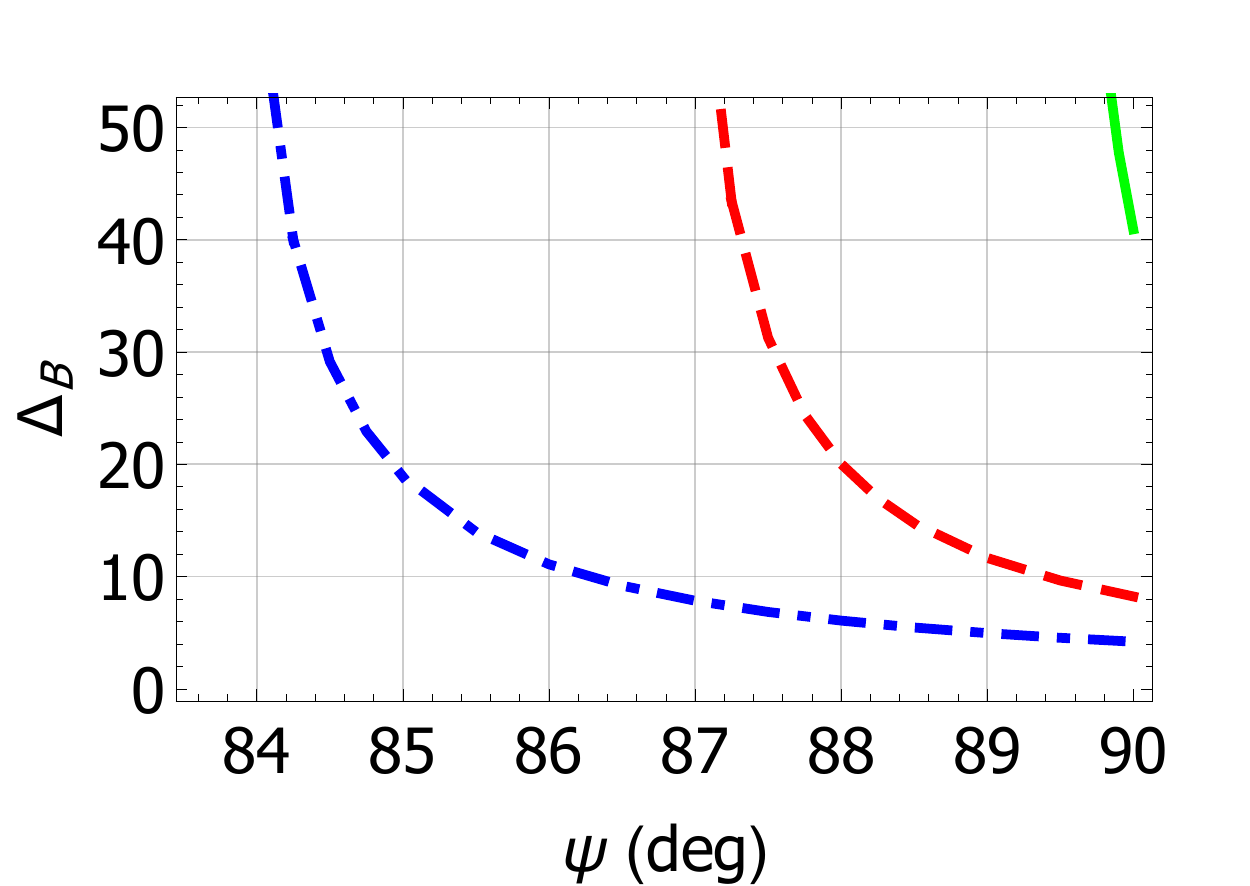}
 \caption{As Fig.~\ref{fig2} but for the first solution branch when $\chi = 45^\circ$.
 } \label{fig4}
\end{figure}

\begin{figure}[!htb]
\centering
\includegraphics[width=4.3cm]{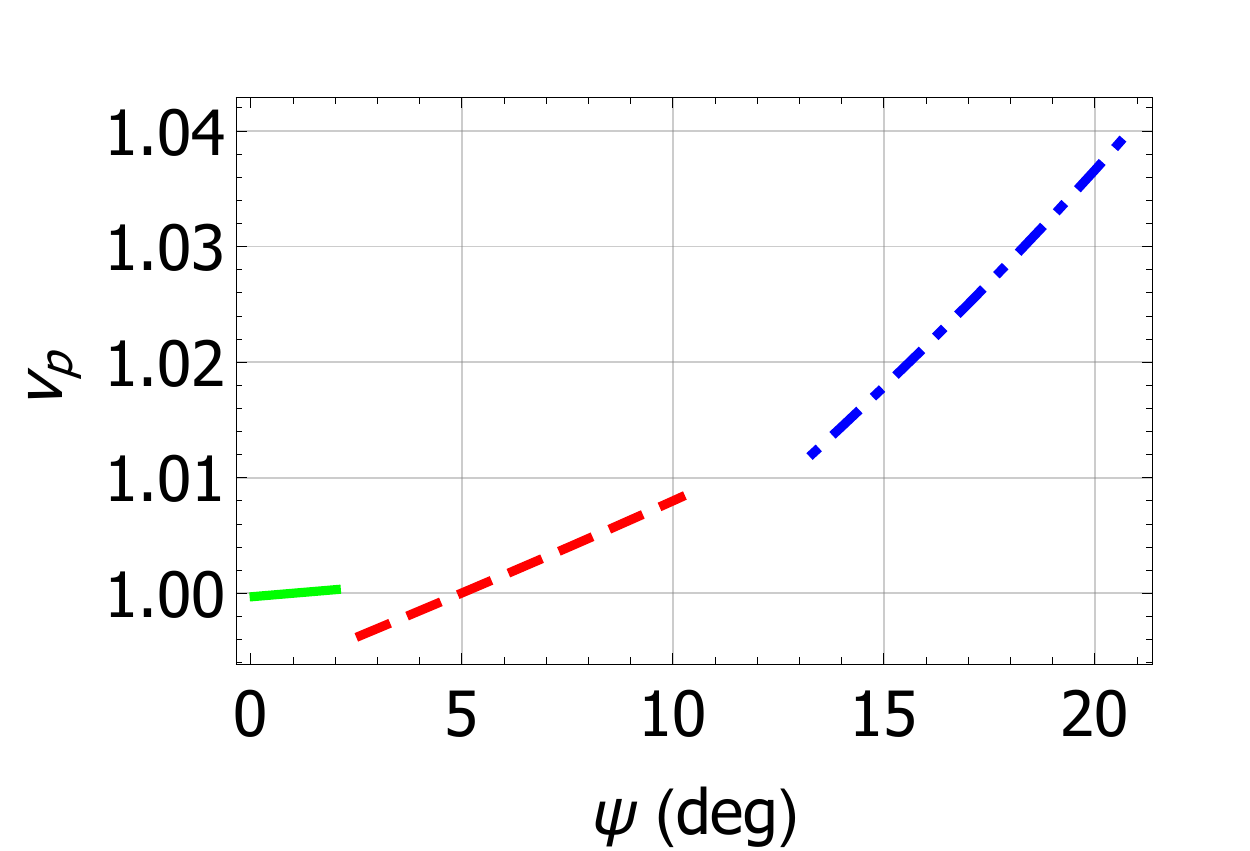}
\includegraphics[width=4.3cm]{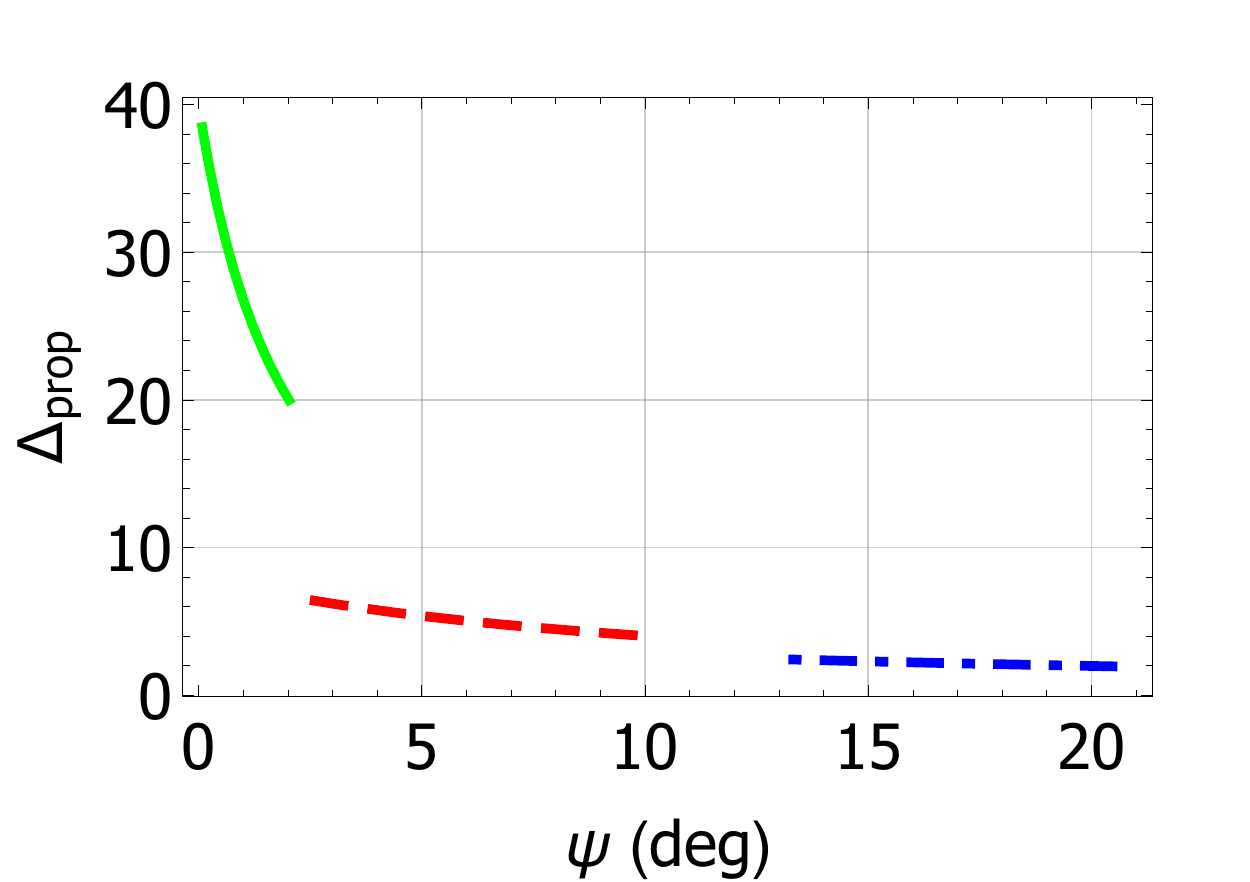}\\
\includegraphics[width=4.3cm]{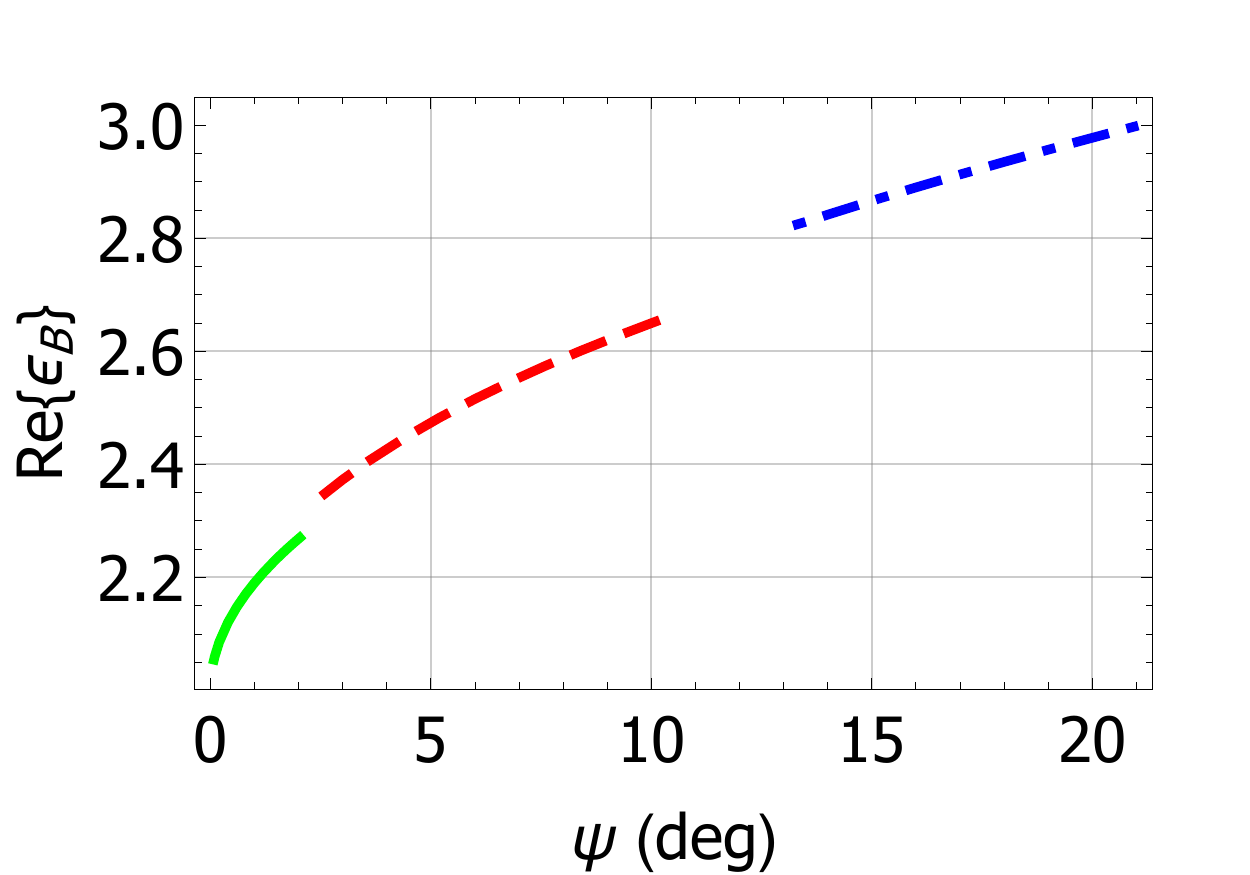}
\includegraphics[width=4.3cm]{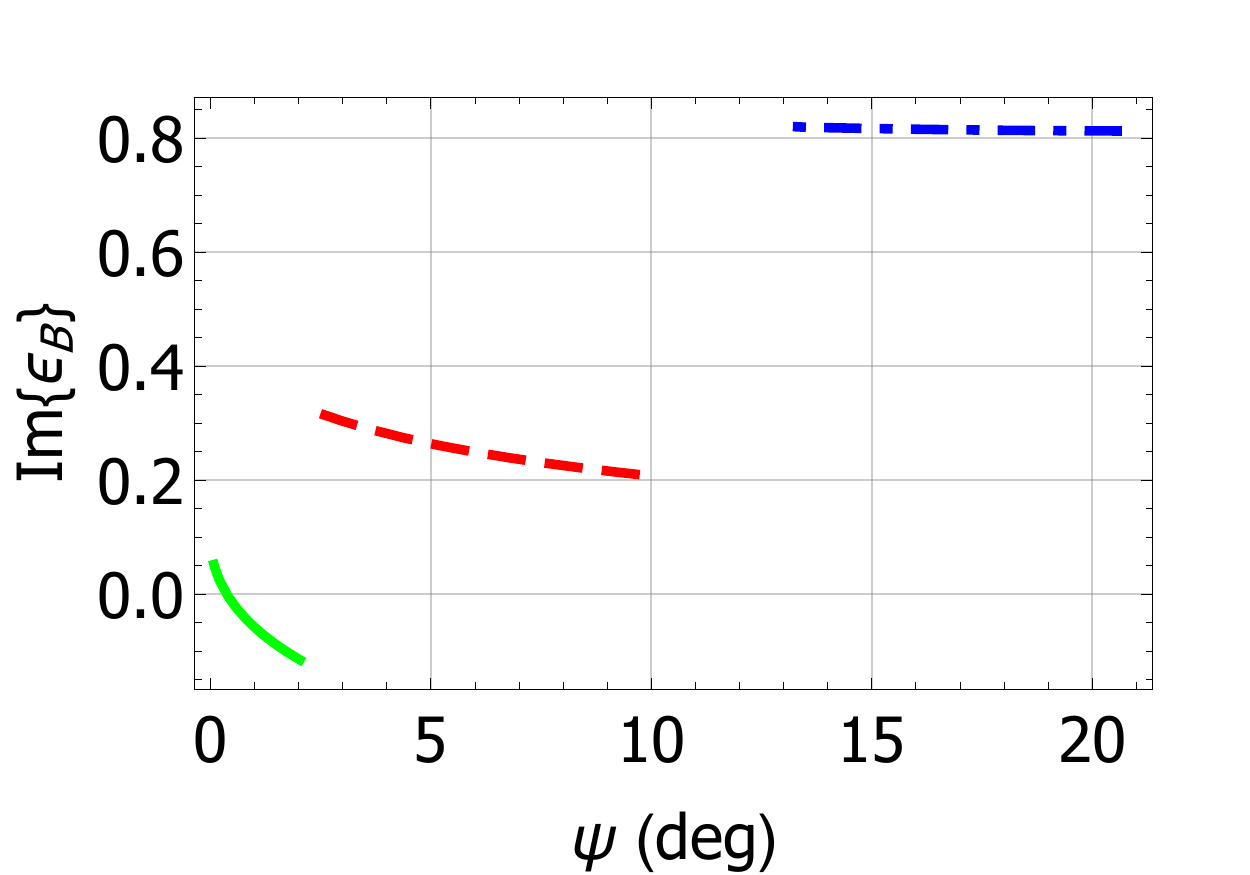}\\
\includegraphics[width=4.3cm]{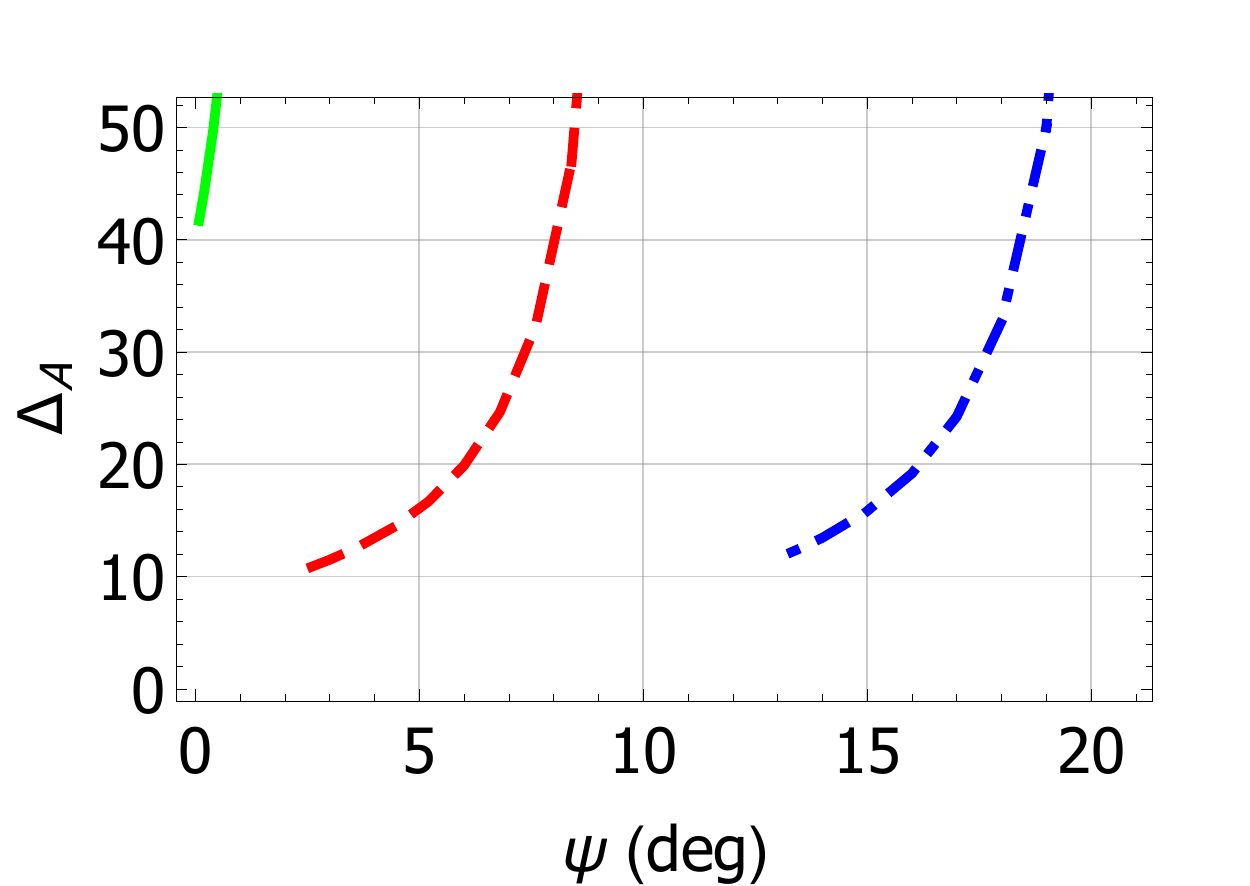}
\includegraphics[width=4.3cm]{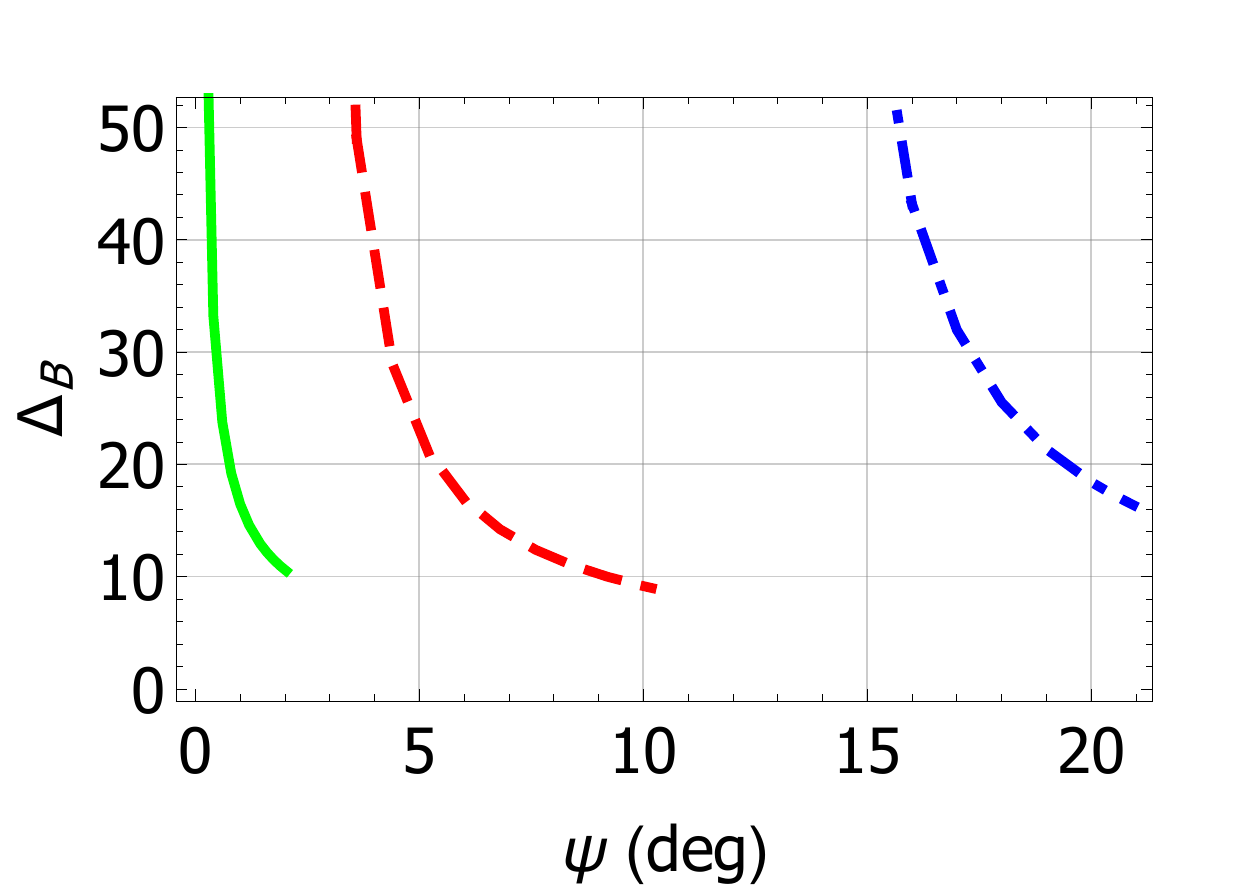}
 \caption{As Fig.~\ref{fig2} but for the second solution branch when $\chi = 45^\circ$.
 } \label{fig5}
\end{figure}

\begin{figure}[!htb]
\centering
\includegraphics[width=4.3cm]{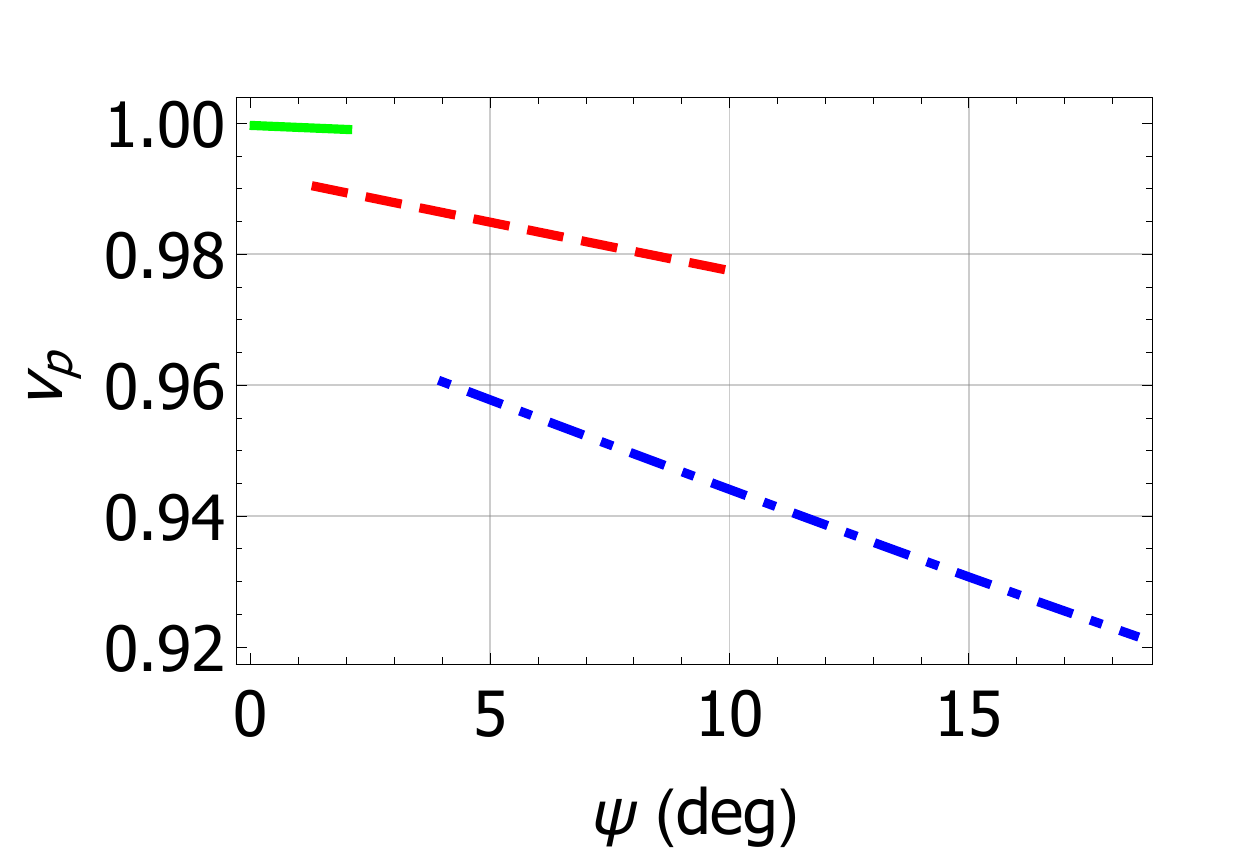}
\includegraphics[width=4.3cm]{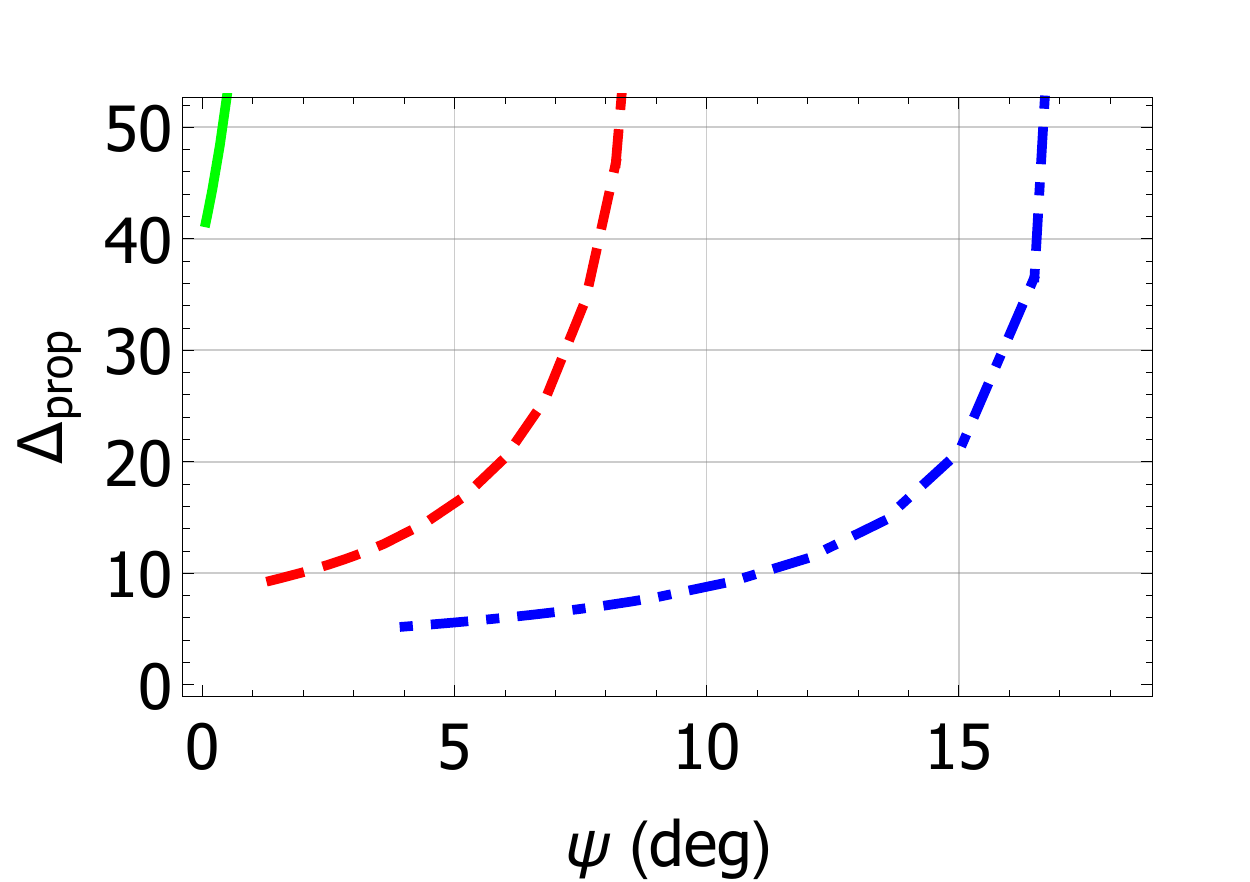}\\
\includegraphics[width=4.3cm]{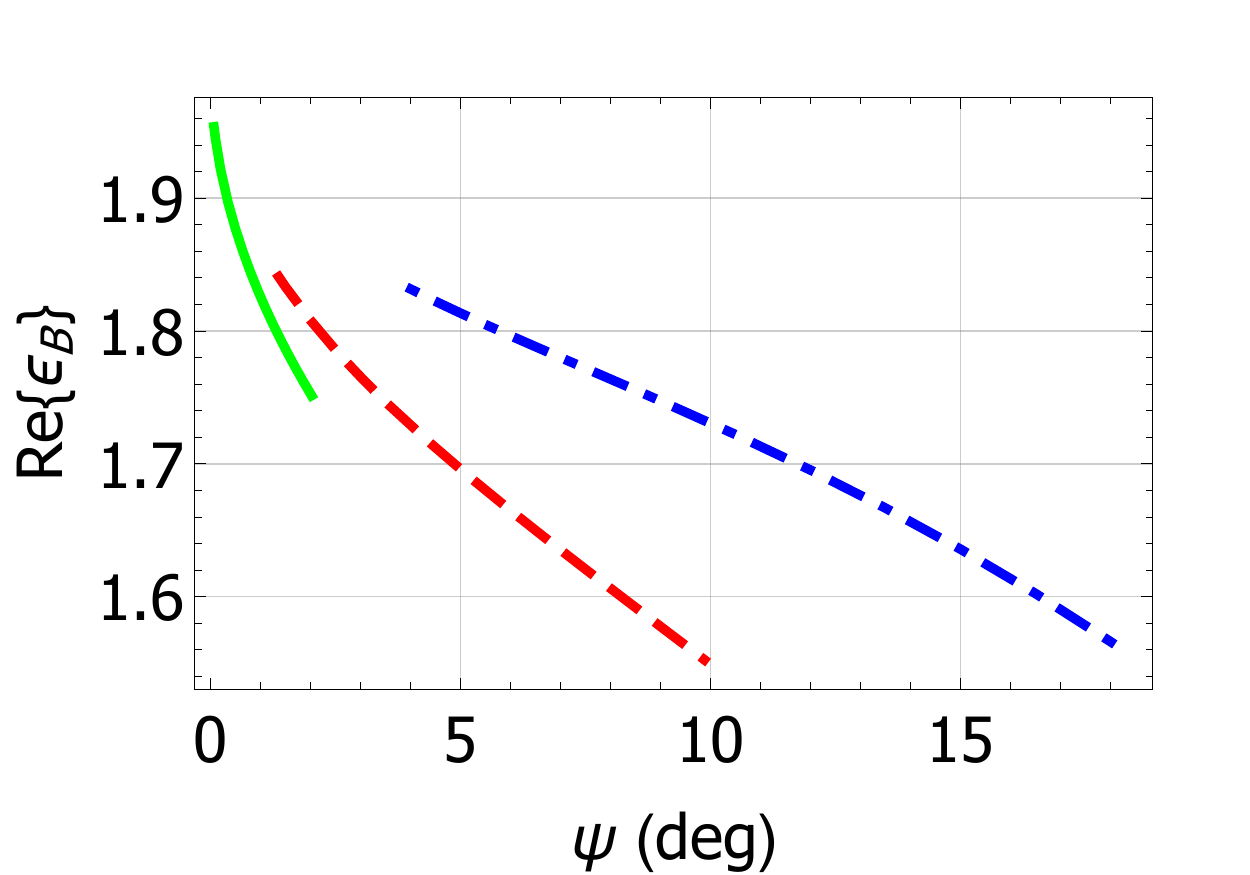}
\includegraphics[width=4.3cm]{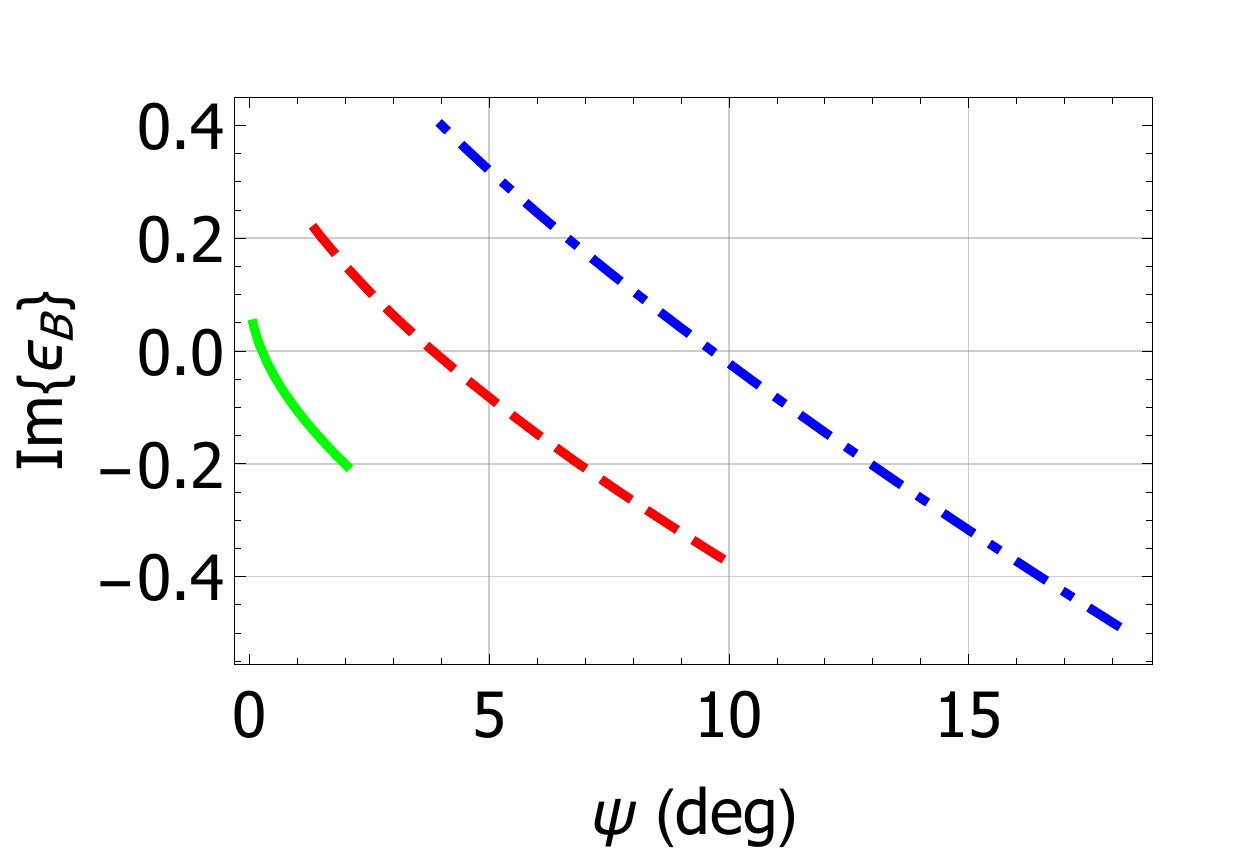}\\
\includegraphics[width=4.3cm]{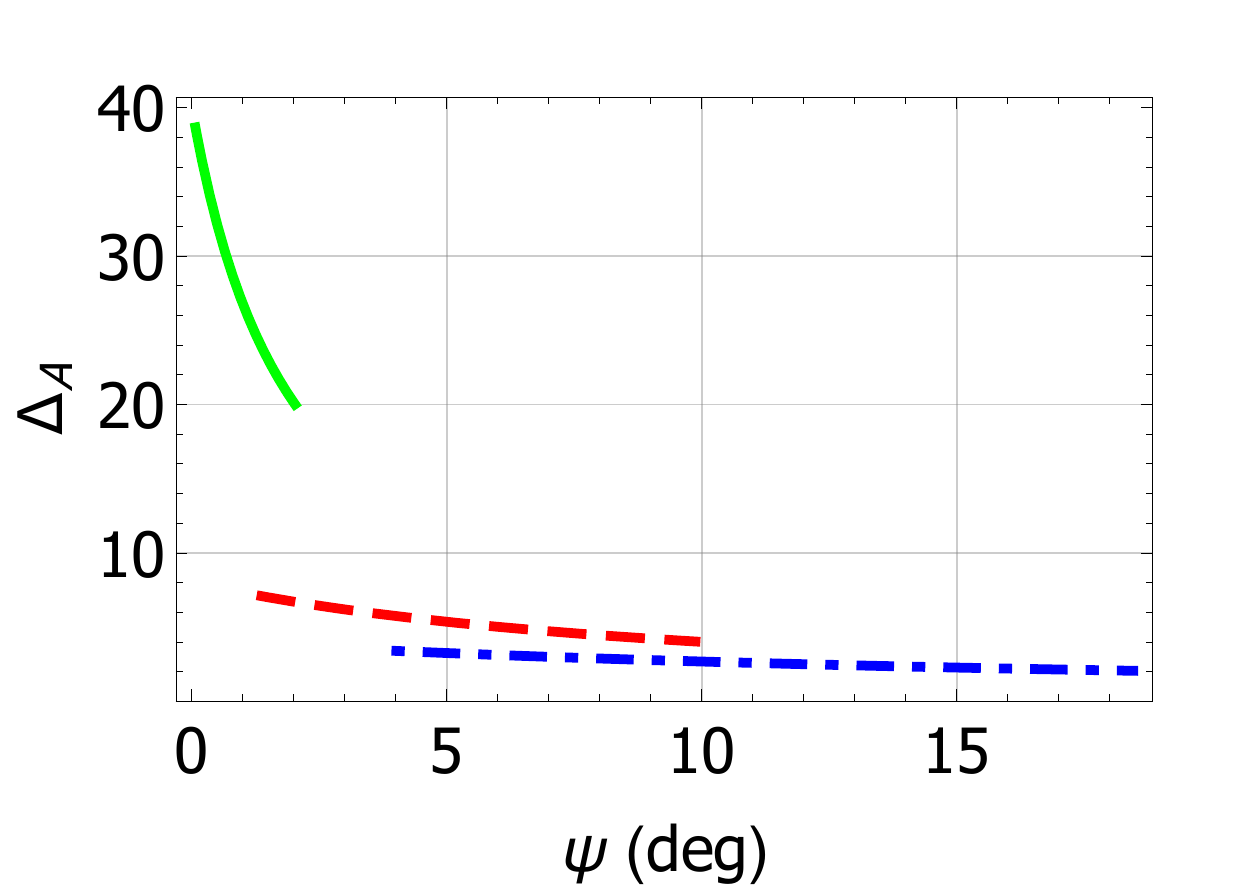}
\includegraphics[width=4.3cm]{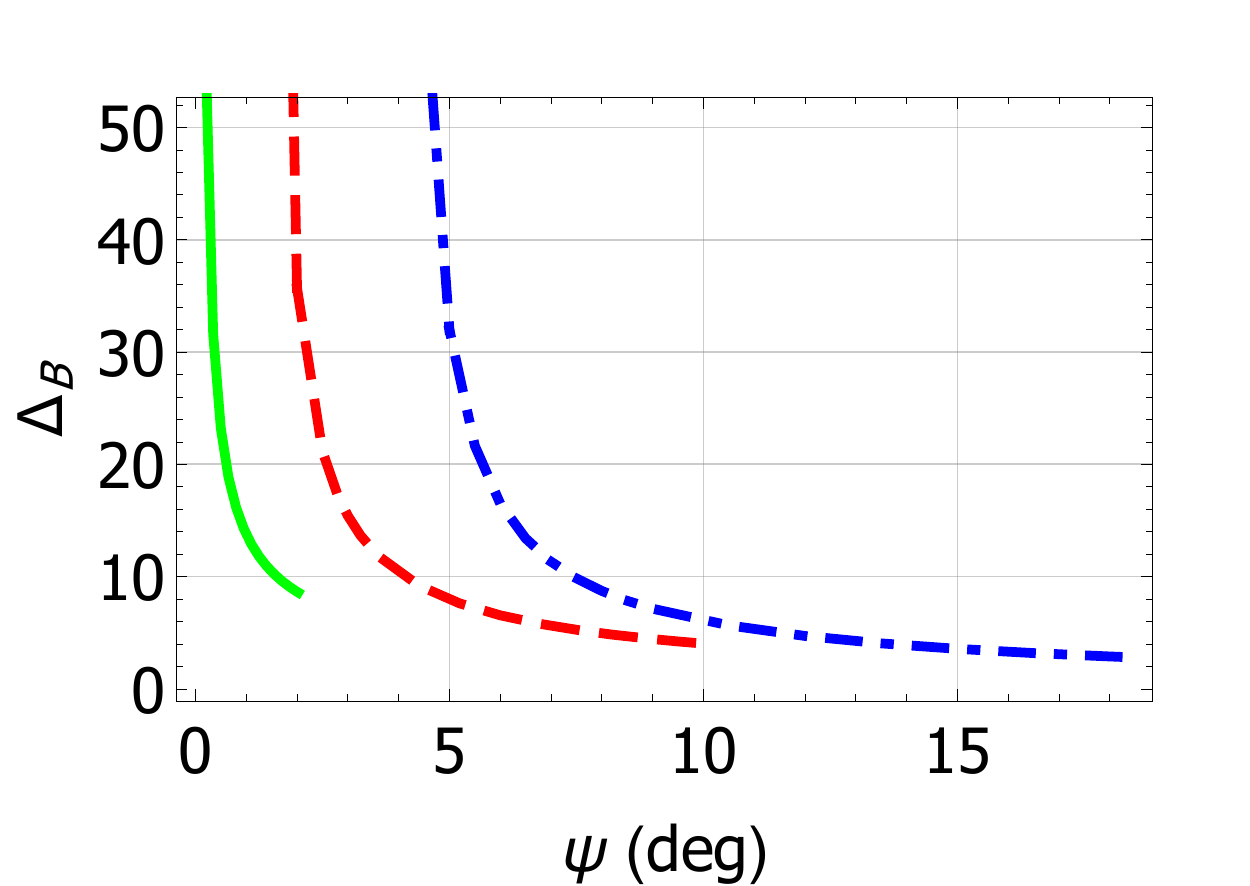}
 \caption{As Fig.~\ref{fig2} but for the third solution branch when $\chi = 45^\circ$.
 } \label{fig6}
\end{figure}

For the sake of illustration, we focus our attention on 
$\chi = 45^\circ$, which is equidistant from both extremities of the range $0^\circ<\chi<90^\circ$.
Zero, one, or two solutions of the dispersion  equation \r{dispersion_eq}
can be found for each value of $\delta$, depending on the
value of $\psi$. These solutions can be organized into three branches with $\psi$ as a variable.
Whereas the first  branch spans only   large values of $\psi$, as becomes clear from Fig.~\ref{fig4},
Figs.~\ref{fig5} and \ref{fig6} show that the second and the third branches span only small values of $\psi$, respectively. The $\psi$-range in which at least one solution exists is of quite small extent
($\sim 2^\circ$) when $\delta=0.1$, but that extent widens to $\sim 9^\circ$ when $\delta=1$.

For the first solution branch, as illustrated in Fig.~\ref{fig4}, 
the dispersion  equation \r{dispersion_eq} yields values of $\eps_\calB$ for $\psi$ close to $90^\circ$. As $\mbox{Re} \lec \eps_\calB \ric < 0 $
and  $\mbox{Im} \lec \eps_\calB \ric < 0 $ for all values of $\delta$, the partnering material $\calB$ must be an active plasmonic material
\c{active_plasmonic,active_plasmonic2}.
The magnitudes of both  $\mbox{Re} \lec \eps_\calB \ric  $
and  $\mbox{Im} \lec \eps_\calB \ric  $ are larger for larger values of $\delta$.
The magnitudes of the phase speeds in Fig.~\ref{fig4} are much larger than the corresponding phase speeds in Fig.~\ref{fig2}. Also, 
$v_{\text{p}} < 0 $ in Fig.~\ref{fig4}  for all values of $\delta$, whereas 
$v_{\text{p}} > 0 $  for all values of $\delta$ in Fig.~\ref{fig2}. Thus, the
DV  surface waves 
on the  first solution branch
propagate with negative phase velocity; this phenomenon has recently been reported for surface waves supported by hyperbolic materials \c{ML_hyperbolic_NPV}.
At  a fixed value of $\psi$, the propagation length $\Delta_{\text{prop}}$, and the penetration depths $\Delta_\calA$ and $\Delta_\calB$, in
Fig.~\ref{fig4} are all larger when $\delta$ is smaller. In addition, $\Delta_\calB$ becomes unbounded as $\psi$ approaches its lowest value.

For the second solution branch,  $\mbox{Re} \lec \eps_\calB \ric > 0 $ for all values of $\delta$
  in Fig.~\ref{fig5}. Also
  $\mbox{Im} \lec \eps_\calB \ric  >0$  except for a small range of $\psi$ values  when $\delta = 0.1$.
Thus, for most values of $\delta$ and $\psi$, the partnering material $\calB$ is a dissipative dielectric material.
The magnitudes of both  $\mbox{Re} \lec \eps_\calB \ric  $
and  $\mbox{Im} \lec \eps_\calB \ric  $ are larger for larger values of $\delta$.
The phase speeds in Fig.~\ref{fig5} are much smaller in magnitude than those  in Fig.~\ref{fig4}. Also, all the phase speeds in Fig.~\ref{fig5} are positive, unlike those in Fig.~\ref{fig4}.
For all values of $\delta$ in Fig.~\ref{fig5}, the   penetration depth $\Delta_\calA$    becomes unbounded as $\psi$ approaches its maximum value, whereas the penetration depth $\Delta_\calB$    becomes unbounded as $\psi$ approaches its minimum value.

For the third solution branch,  $\mbox{Re} \lec \eps_\calB \ric > 0 $ for all values of $\delta$
  in Fig.~\ref{fig6}; however, $\mbox{Im} \lec \eps_\calB \ric > 0 $ for small values of $\psi$ 
   but   $\mbox{Im} \lec \eps_\calB \ric < 0 $ for large values of $\psi$. Thus, the partnering material $\calB$ is dissipative dielectric   for small values of $\psi$ and an active dielectric   for large values of $\psi$.
The phase speeds in Fig.~\ref{fig6} are positive, like those in Fig.~\ref{fig5}.
For all values of $\delta$, the  propagation length $\Delta_{\text{prop}}$ 
 in Fig.~\ref{fig6} 
 becomes unbounded as $\psi$ approaches its maximum value, while the penetration depth $\Delta_\calB$  becomes unbounded as $\psi$ approaches its minimum value.

\begin{figure}[!htb]
\centering
\includegraphics[width=4.3cm]{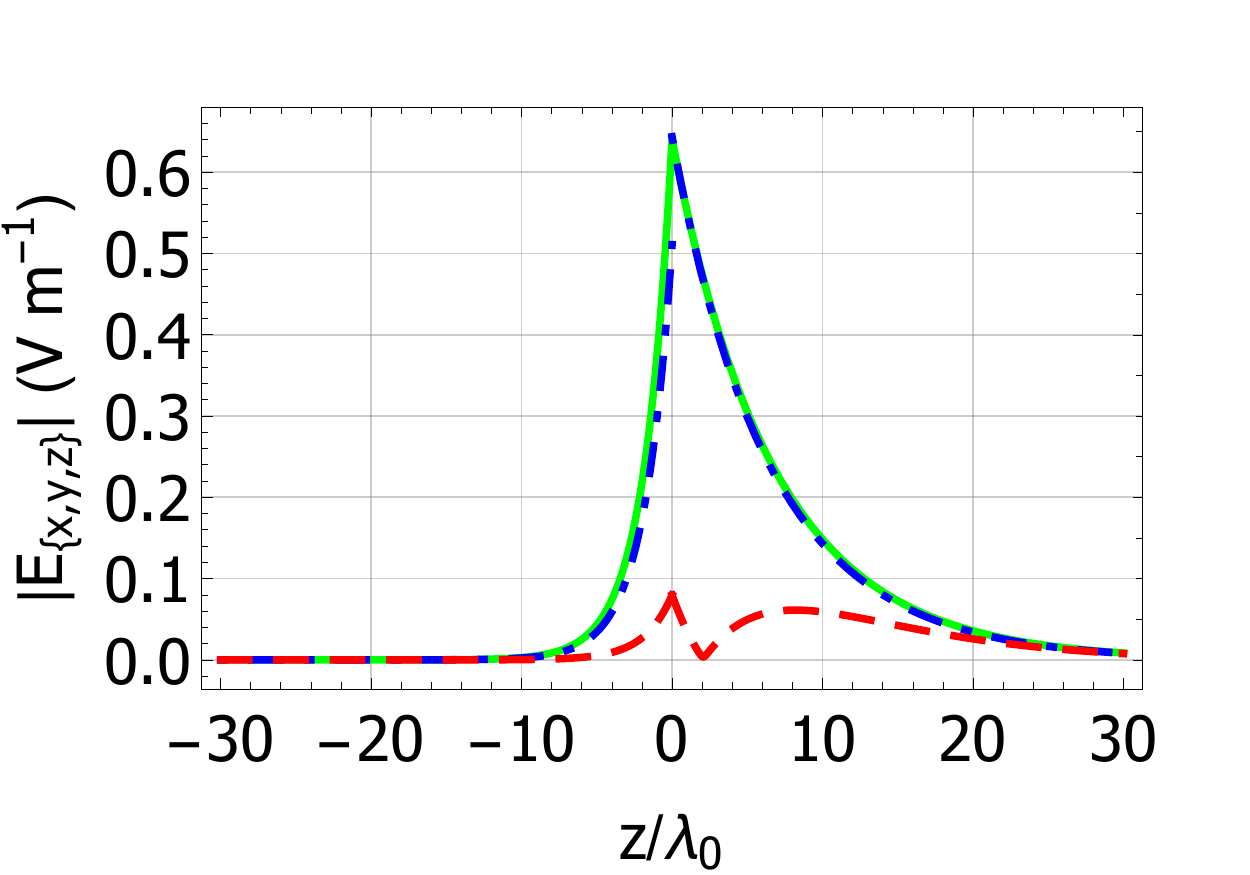}
\includegraphics[width=4.3cm]{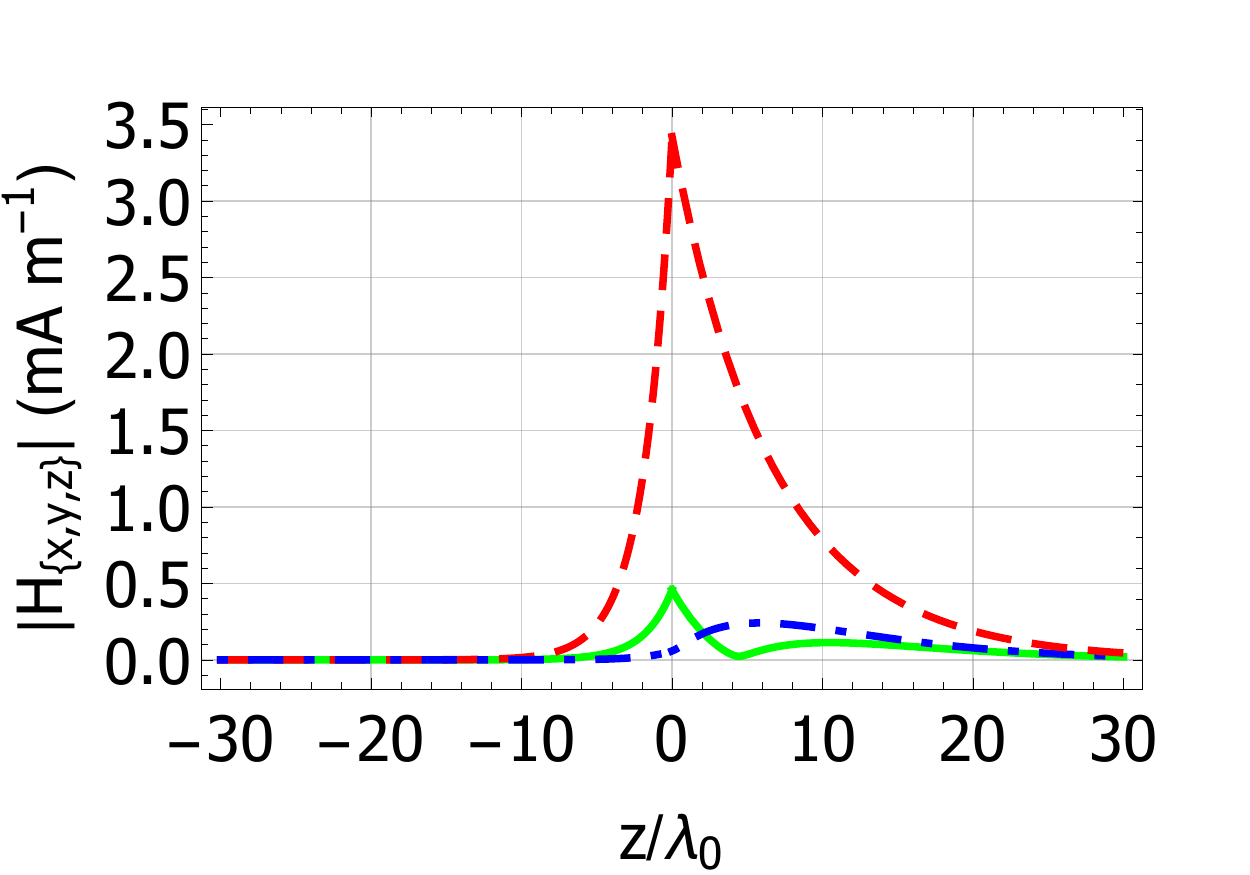}\\
\includegraphics[width=4.3cm]{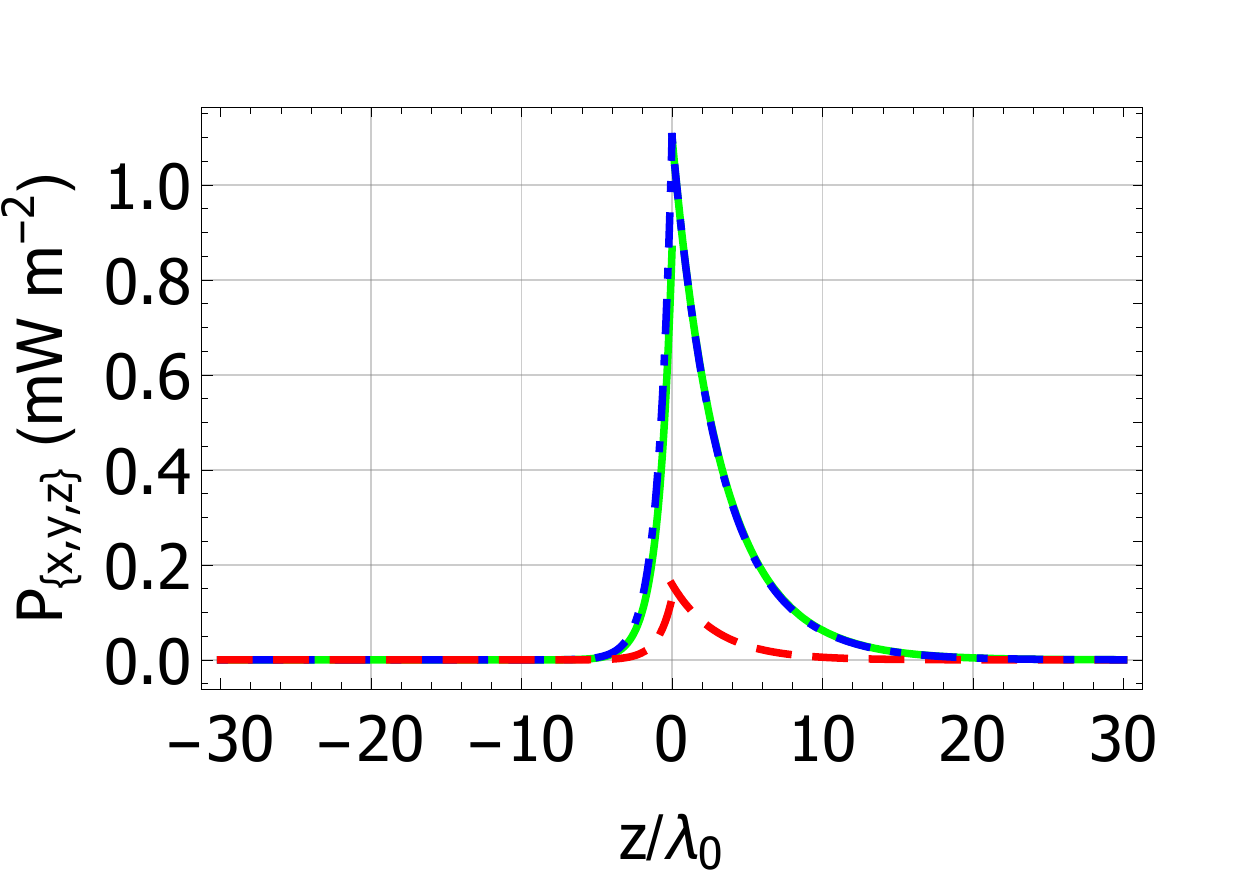}
 \caption{As Fig.~\ref{fig3} but 
  for $\delta=0.5$, 
 $\chi = 45^\circ$, and $\psi = 8^\circ$, so that  $\eps_\mathcal{B} = 2.588+0.266i$ for  DV surface-wave propagation.
 } \label{fig7}
\end{figure}

In Fig.~\ref{fig7}, the
spatial profiles of the magnitudes of the  Cartesian components
of $\#E (z \uz)$ and $\#H (z\uz)$, as well as the Cartesian components
of
$\#P (z\uz)$,
are presented 
for a solution lying on the   branch depicted in   Fig.~\ref{fig5} for $\delta=0.5$.  
  The relative rate of decay of the fields in material $\calB$ as $|z|/\lambdao$ increases is much faster in 
  Fig.~\ref{fig7} than   in Fig.~\ref{fig3}. Also, 
there is relatively more
energy flow  concentrated in directions normal to the interface plane $z=0$
in Fig.~\ref{fig7}  than   in Fig.~\ref{fig3}.

\subsubsection{$\chi = 85^\circ$ \label{sec3B2}}

\begin{figure}[!htb]
\centering
\includegraphics[width=4.3cm]{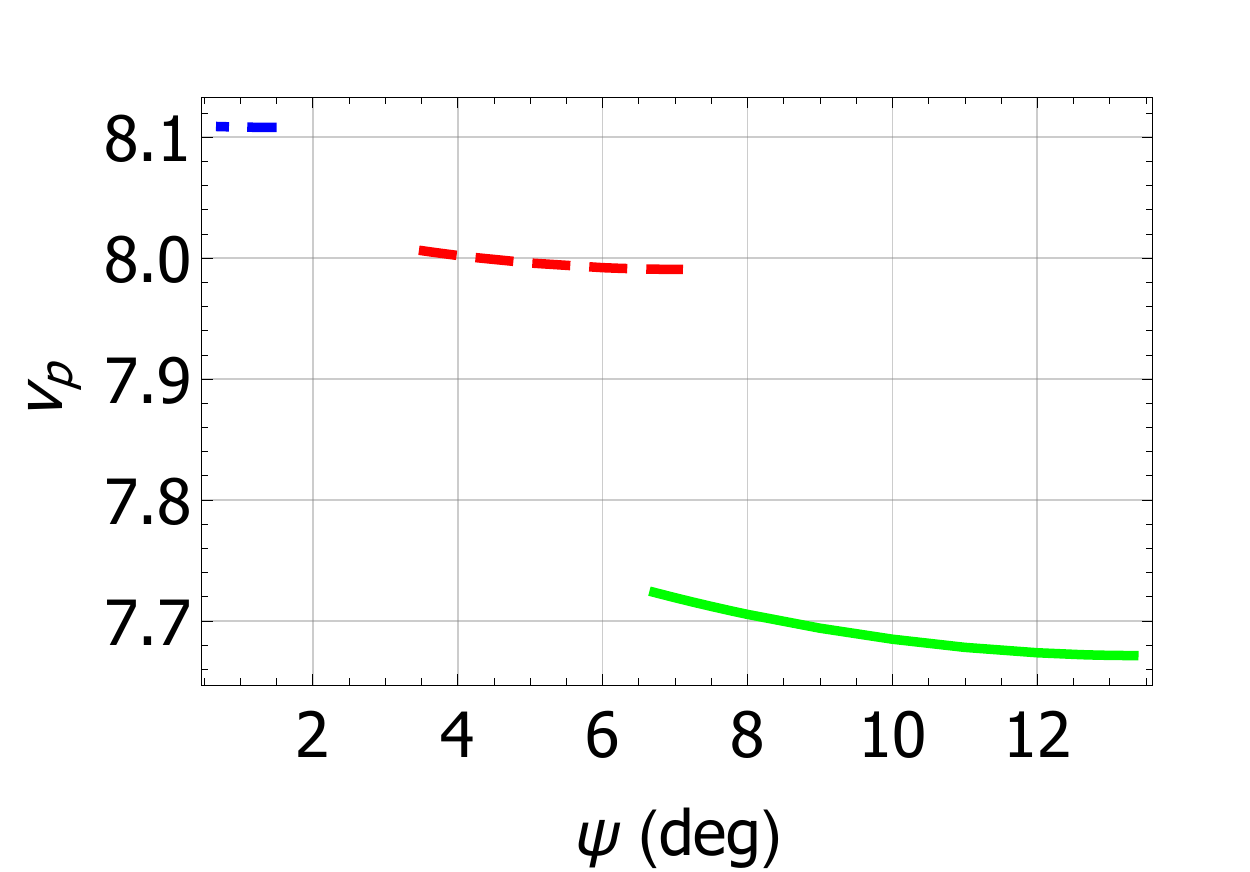}
\includegraphics[width=4.3cm]{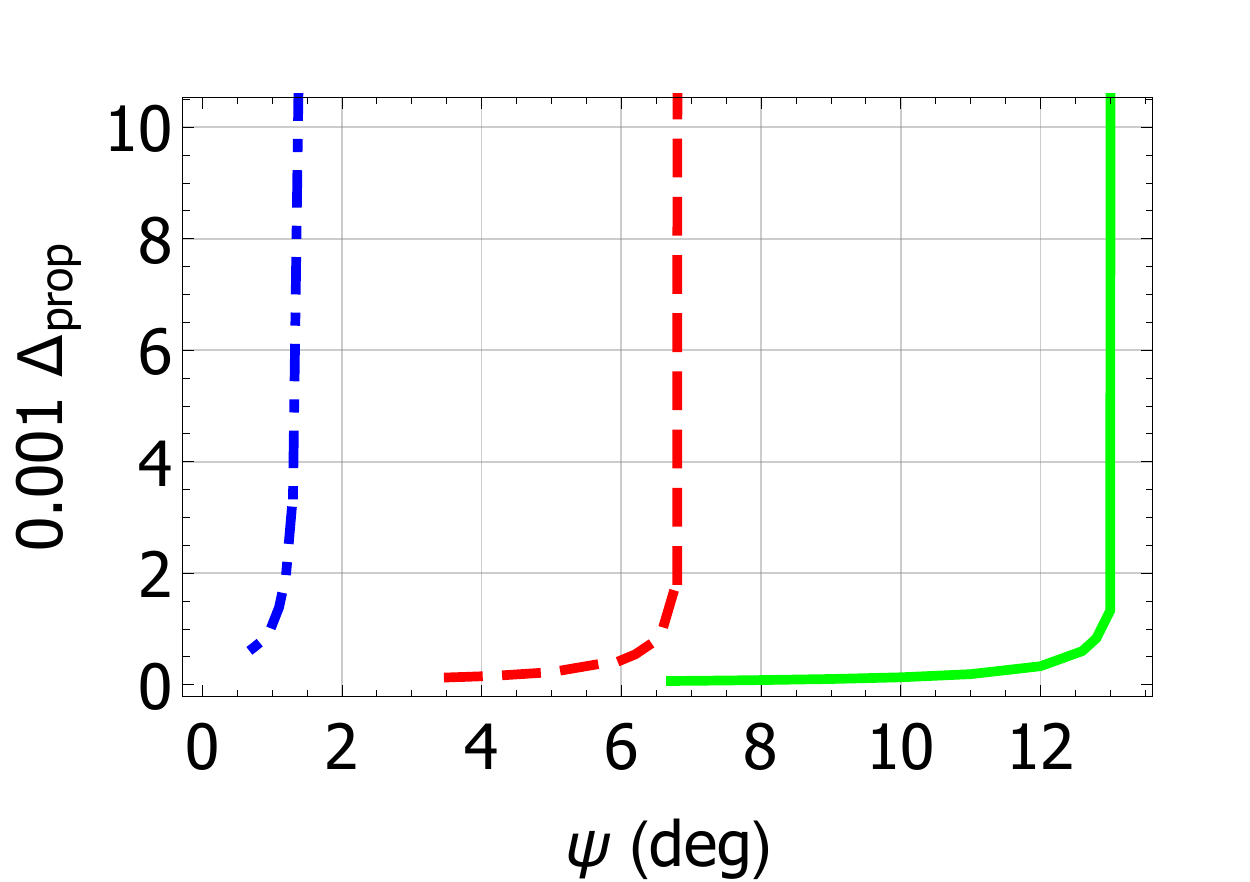}\\
\includegraphics[width=4.3cm]{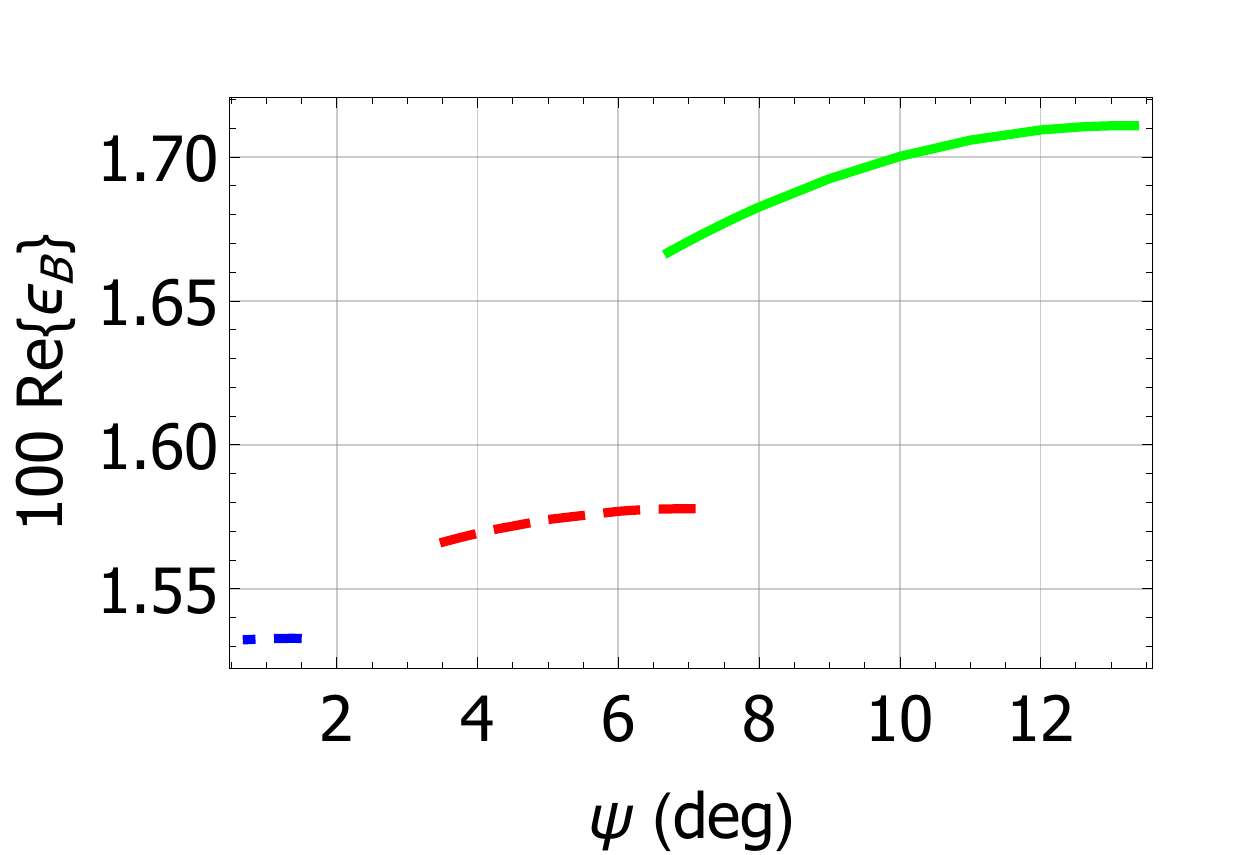}
\includegraphics[width=4.3cm]{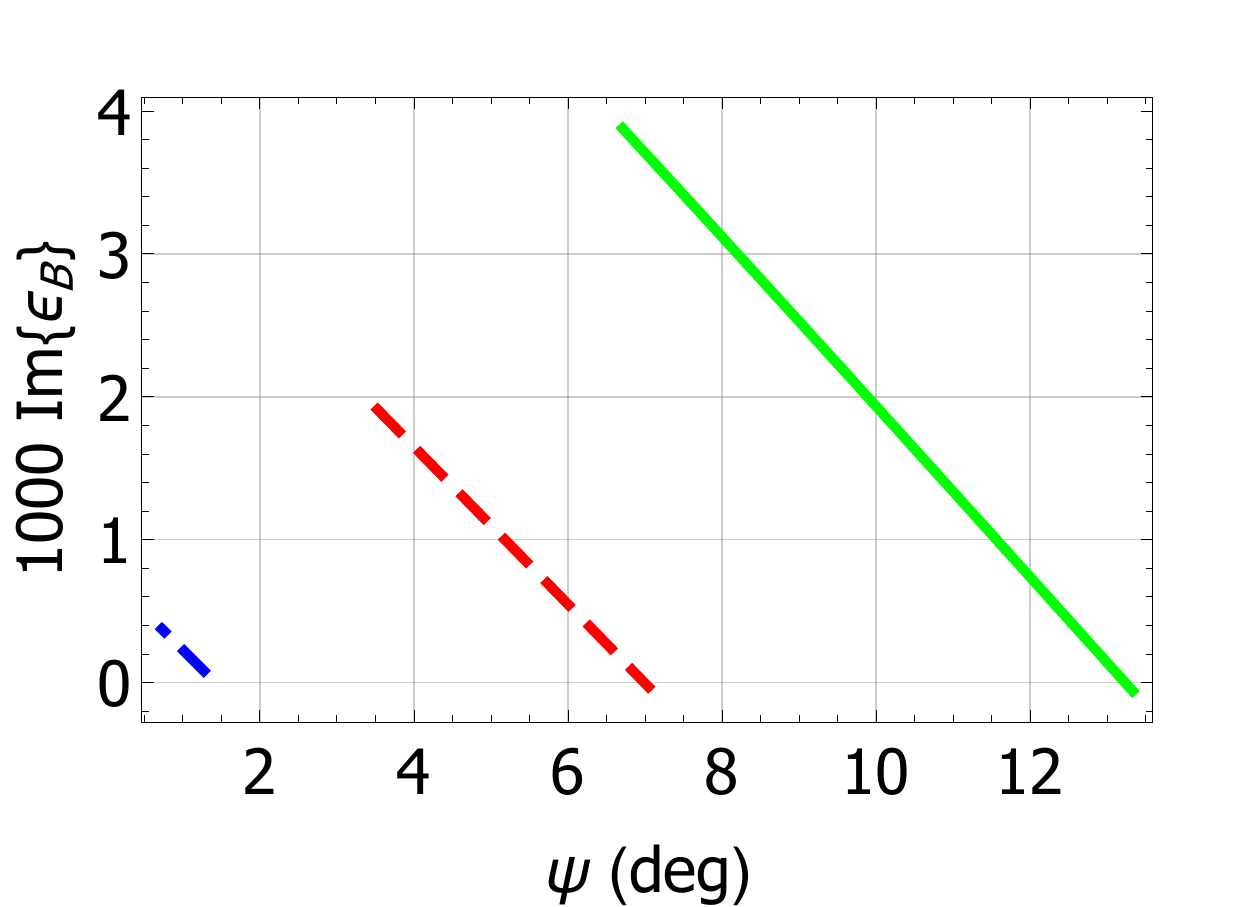}\\
\includegraphics[width=4.3cm]{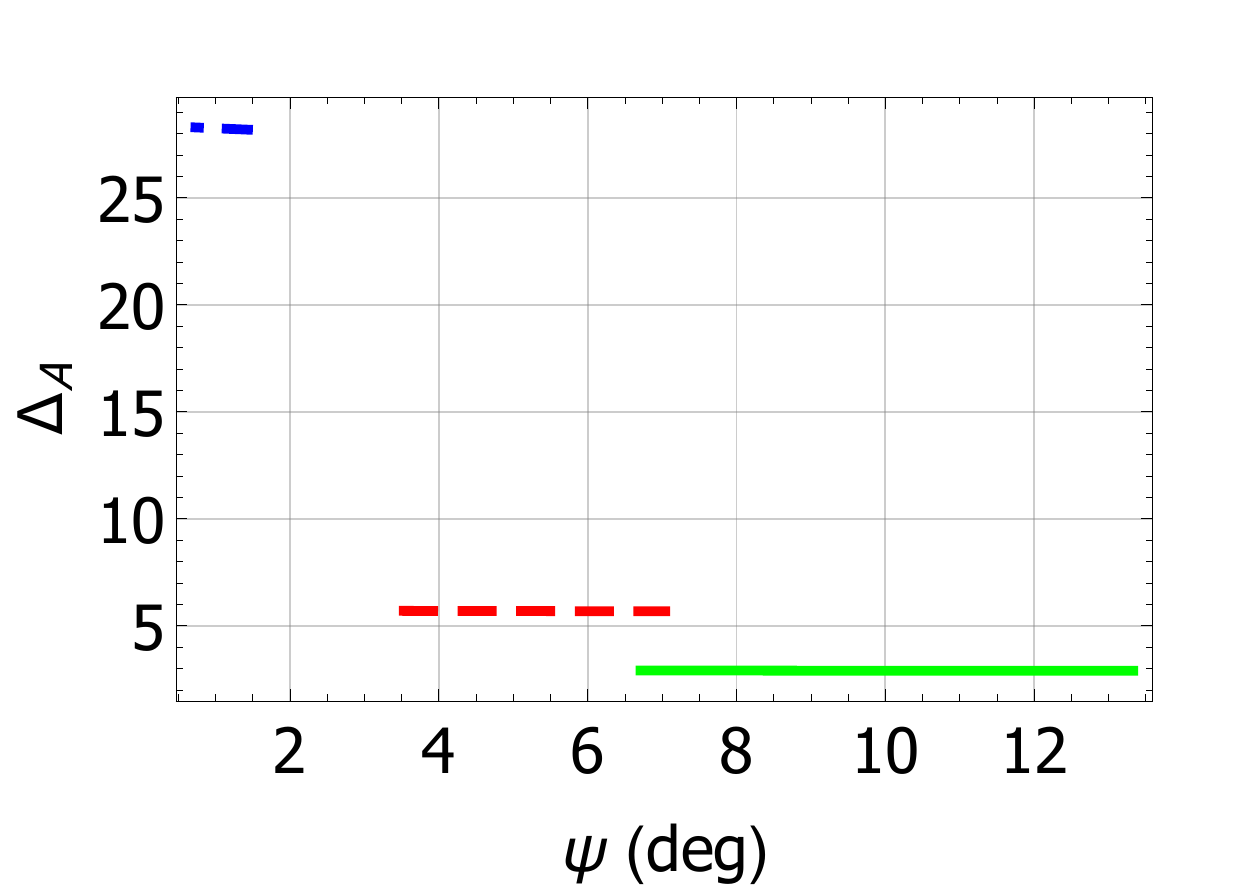}
\includegraphics[width=4.3cm]{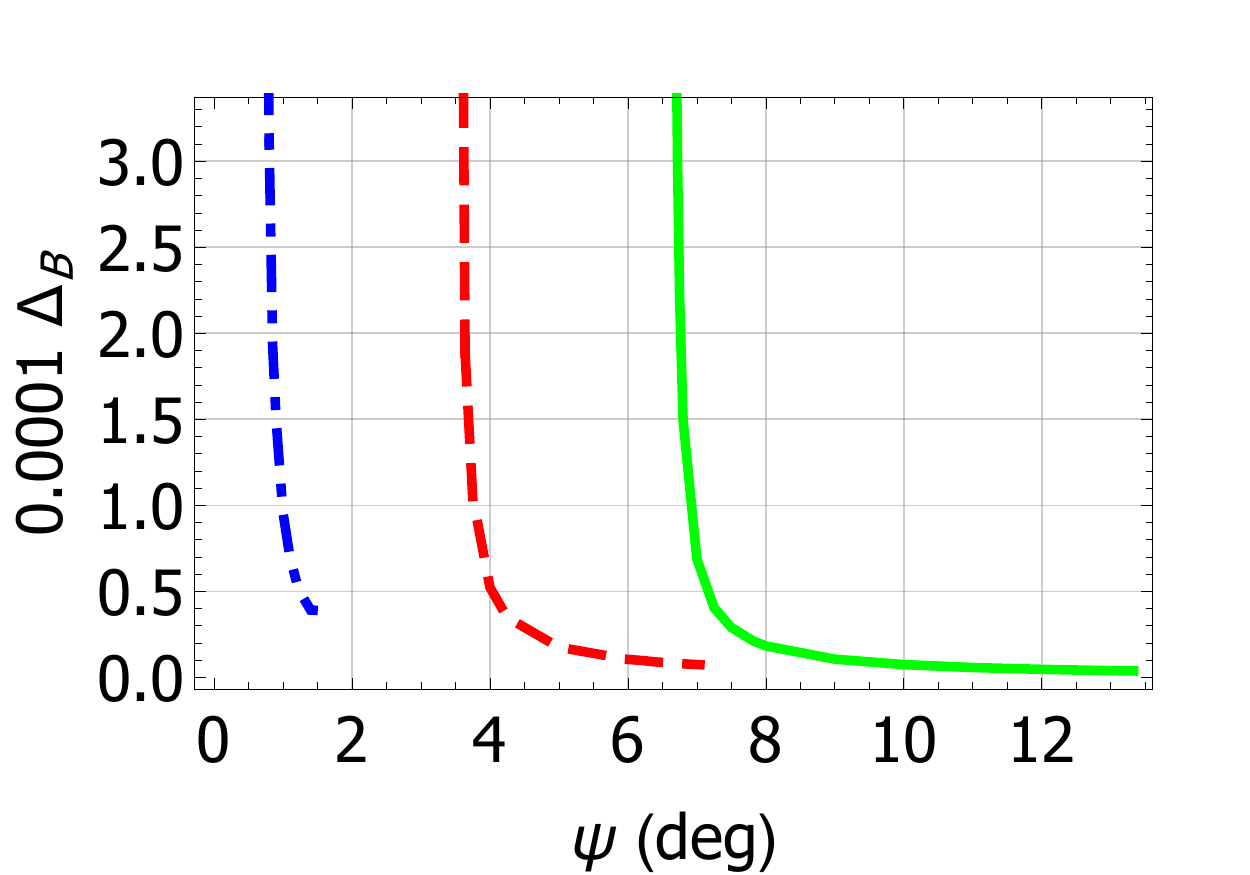}
 \caption{As Fig.~\ref{fig2} but  when $\chi = 85^\circ$.
 } \label{fig8}
\end{figure}

As DV surface-wave propagation is not possible for $\psi=90^\circ$, finally
we  explore the  case where $\chi = 85^\circ$.
Unlike the three solution branches shown
for $\chi = 45^\circ$ in Sec.~3.\ref{sec3B1}, only one solution
branch exists for $\chi = 85^\circ$, regardless of the value of $\delta>0$.
In Fig.~\ref{fig8},
  $\mbox{Re} \lec \eps_\calB \ric > 0 $
and  $\mbox{Im} \lec \eps_\calB \ric > 0 $ for all values of $\delta$.
The magnitudes of both the real and imaginary parts of $\eps_\calB$ are  tiny compared to those  for $\chi=0^\circ$
(Fig.~\ref{fig2}) and $\chi=45^\circ$ (Figs.~\ref{fig4}--\ref{fig6}). 
Additionally, $\mbox{Im} \lec \eps_\calB \ric$ is approximately one order of magnitude smaller than $\mbox{Re} \lec \eps_\calB \ric$.
Thus, material $\calB$ is a  dissipative dielectric material. 
 Also,  $\mbox{Re} \lec \eps_\calB \ric  $ increases, but $\mbox{Im} \lec \eps_\calB \ric  $
 decreases, as $\psi$ increases.

 The phase speeds presented in Fig.~\ref{fig8} are 
positive  and
relatively large, with $v_{\text{p}}$ being larger  when $\delta$ is larger.
The  propagation lengths $\Delta_{\text{prop}}$ and the penetration depths $\Delta_\calB$ presented in Fig.~\ref{fig8} are generally tiny, but $\Delta_{\text{prop}}$  rapidly becomes unbounded as $\psi$ approaches its maximum value and $\Delta_\calB$  rapidly becomes unbounded as $\psi$ approaches its minimum value.

\begin{figure}[!htb]
\centering
\includegraphics[width=4.3cm]{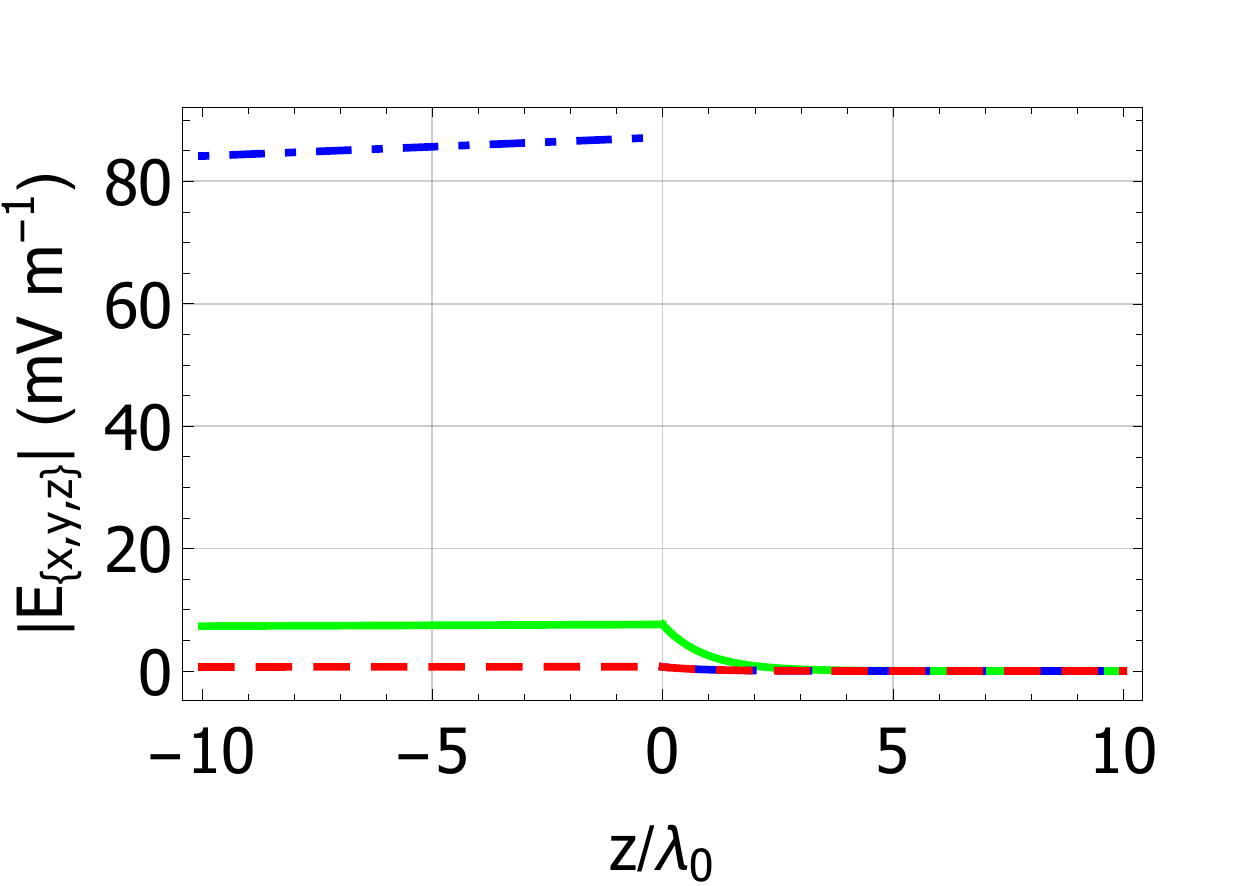}
\includegraphics[width=4.3cm]{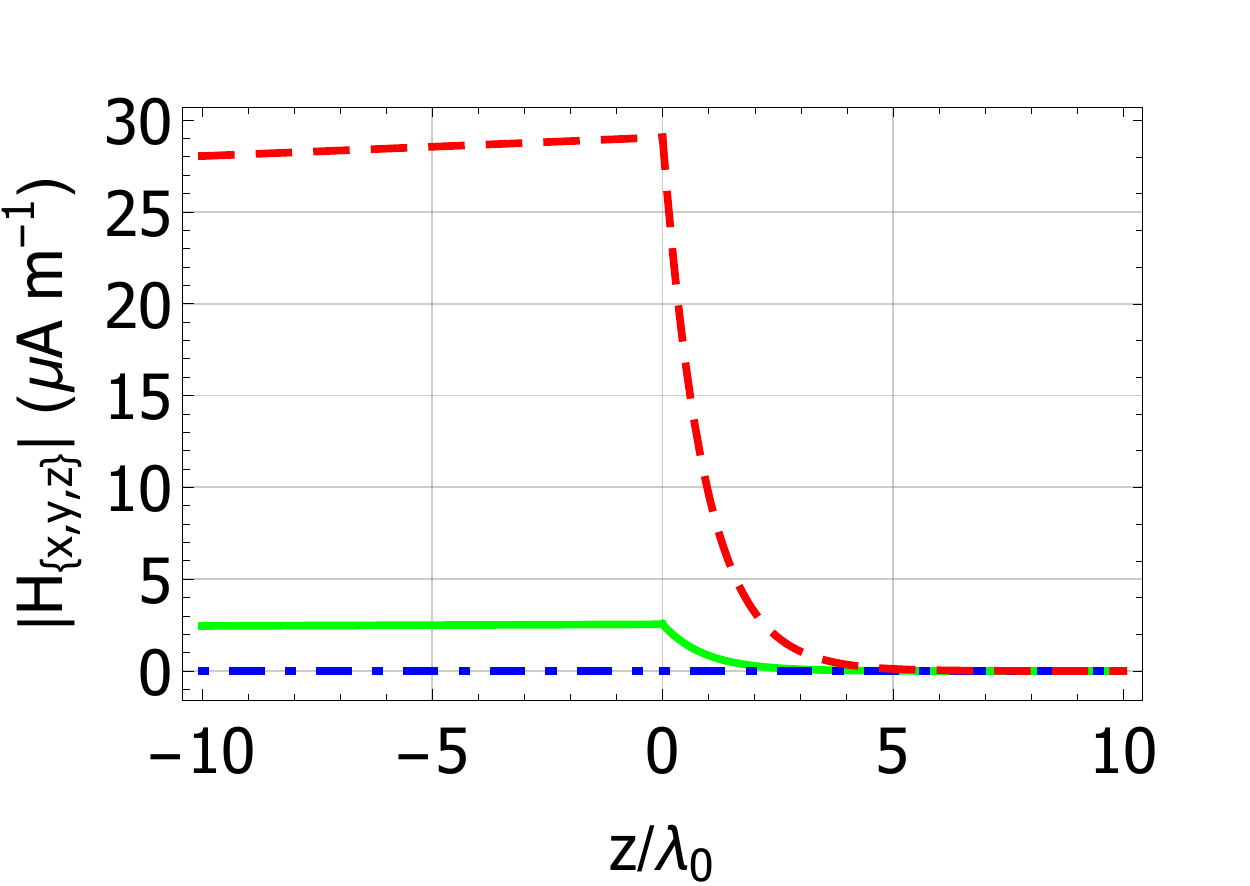}\\
\includegraphics[width=4.3cm]{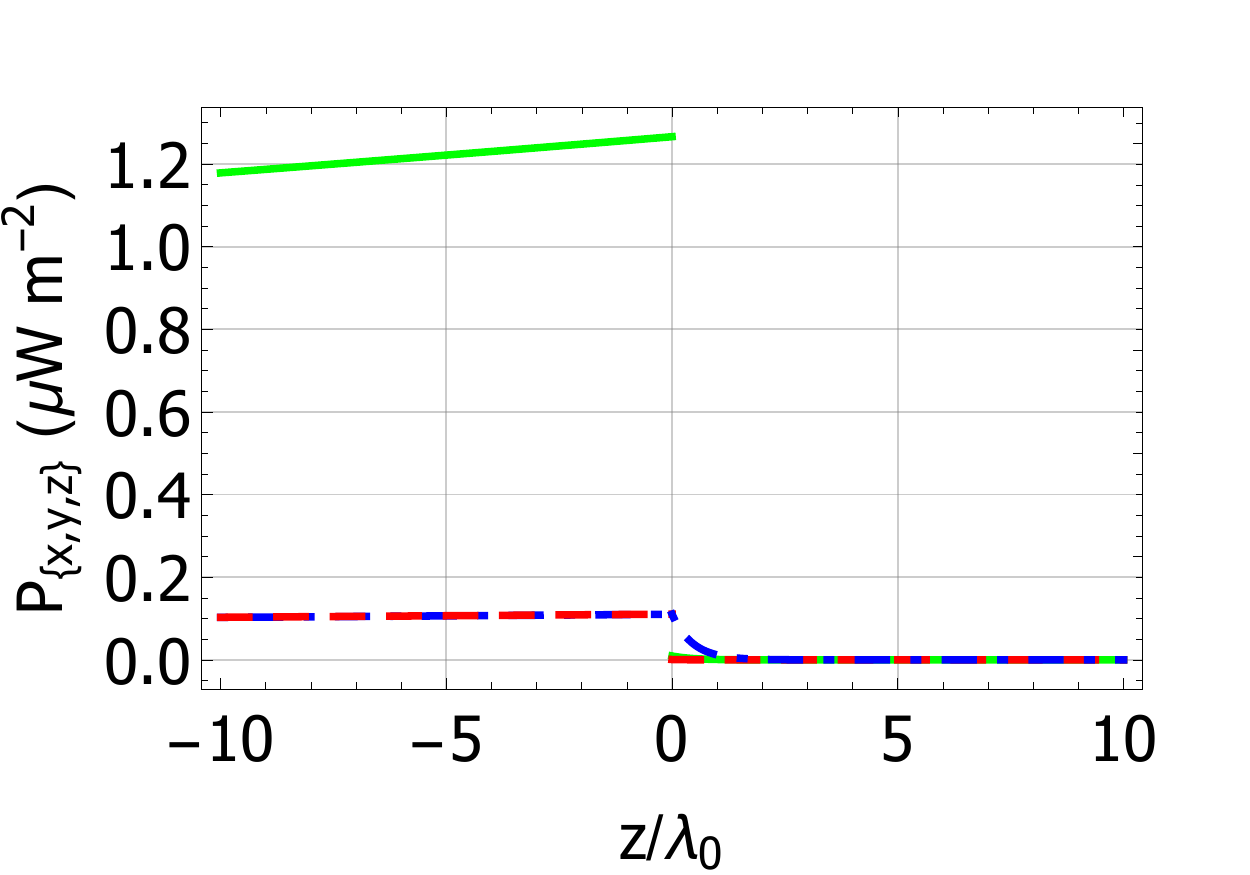}
 \caption{As Fig.~\ref{fig3} but 
  for $\delta=0.5$, 
 $\chi = 85^\circ$, and $\psi = 5^\circ$, so that    $\eps_\mathcal{B} = 0.016+0.001i$ for  DV surface-wave propagation.
 } \label{fig9}
\end{figure}

Spatial profiles of the
  magnitudes of the  Cartesian components
of $\#E (z \uz)$ and $\#H (z\uz)$, as well as spatial profiles of the Cartesian components
of
$\#P (z\uz)$,
are presented in Fig.~\ref{fig9} for
$\delta = 0.5$ and $\psi = 5^\circ$, so that $\eps_\calB = 0.016+0.001i$.
Parethetically, such values of relative permittivity that are close to zero can be realized using metamaterial technologies \c{ENZ1,ENZ2}.
  The plots in Fig.~\ref{fig9} are qualitatively similar to those in Fig.~\ref{fig3},  the most obvious difference being that  
there is relatively more
energy flow  concentrated in directions normal to the interface plane $z=0$, especially in material $\calB$,  than  in Fig.~\ref{fig3}.

\section{Closing remarks}

Dyakonov--Voigt  surface waves guided by the planar interface 
of a uniaxial dielectric material $\calA$ and an isotropic dielectric material $\calB$
  were  numerically investigated by formulating and solving the corresponding canonical boundary-value problem.
  The degree of dissipation and the inclination angle $\chi$ of the optic axis of material $\calA$ 
  were found to profoundly influence  the phase speeds, propagation lengths, and penetration depths of the DV surface waves. Notably, for mid-range values of $\chi$, DV surface waves with negative phase velocity were found when material $\calA$ is dissipative and material $\calB$ is active. For fixed  values
of $\eps^s_\calA$ and $\eps^t_\calA$ in the upper-half-complex plane, DV surface-wave propagation is only possible for large values of $\chi$ when $| \eps_\calB|$ is very small.
  
  In the presented numerical studies, the relative permittivity parameters
  $\eps^s_\calA$ and $\eps^t_\calA$ were fixed (in the upper-half-complex plane) and the dispersion equation
  \r{dispersion_eq} solved to find $\eps_\calB$. 
  We also conducted numerical studies (not presented  here) wherein
    $\eps^t_\calA$ and $\eps_\calB$ were fixed (in the upper-half-complex plane)
  and the dispersion equation
  \r{dispersion_eq} was solved to find $\eps^s_\calA$. Analogously to the presented studies, these other studies 
  revealed that  the imaginary parts of $\eps^t_\calA$ and $\eps_\calB$ have
   a profound influence on the characteristics of the DV surface waves. Furthermore, in this case
   solutions to  the dispersion equation
  \r{dispersion_eq} were found for which $\mbox{Im} \lec \eps^s_\calA \ric < 0$ and $\mbox{Im} \lec \eps^t_\calA \ric > 0$, indicating that material $\calA$ is simultaneously both dissipative and active depending upon propagation direction \c{ML_PRA,ML_Dyakonov}.
  
  Our numerical studies do not reveal physical mechanisms. 
  In the case of Dyakonov surface waves,
  it is well established that dissipative partnering materials support wider angular existence domains   than do
  nondissipative partnering materials
   \c{Sorni,MLieee}.
  However, an analogous
observation is not forthcoming for Dyakonov--Voigt surface waves.

Finally, the numerical studies presented herein were based on
the corresponding canonical boundary-value problem. The canonical 
boundary-value problem represents an idealization: the finite thicknesses of the partnering materials and the means by which the DV surface waves are excited are not taken into account. Nevertheless,
such studies of the canonical 
boundary-value problem
 deliver useful insights into the essential physics of surface-wave propagation, and  represent an essential  step towards the elucidation of 
 DV surface waves in experimental scenarios.

\vspace{5mm}

\noindent {\bf Acknowledgments.}
This work was supported in part by
EPSRC (grant number EP/S00033X/1) and US NSF (grant number DMS-1619901).
AL thanks the Charles Godfrey Binder Endowment at the Pennsylvania State University  and  the Otto M{\o}nsted Foundation for partial support of his research endeavors.

\end{document}